\documentclass[
reprint,
amsmath,amssymb,
aps,
prc,
]{revtex4-2}

\usepackage{mathrsfs}
\usepackage{siunitx}
\usepackage[dvipsnames]{xcolor}
\usepackage{orcidlink}
\usepackage[T1]{fontenc}
\usepackage{graphicx}
\usepackage{dcolumn}
\usepackage{bm}
\usepackage{hyperref}
\hypersetup{
    colorlinks=true,
    linkcolor=blue,
    citecolor=blue,
    filecolor=blue,
    urlcolor=blue,
    pdfborder={0 0 0}
}

\begin{document}

\preprint{APS/123-QED}

\title{Systematic Cranked Shell Model Calculations for \texorpdfstring{$^{87, 89, 91}\text{Br}$}{87, 89, 91 Br}}

\author{Nabeel Salim\orcidlink{0009-0007-4820-1855}}
 \email{nabeel\_s@ph.iitr.ac.in}
\author{Mehak Narula\orcidlink{0009-0003-4746-8401}}
 \email{mehak@ph.iitr.ac.in}
 \author{P. Arumugam\orcidlink{0000-0001-9624-8024}}
 \email{arumugam@ph.iitr.ac.in}
\affiliation{Department of Physics, Indian Institute of Technology Roorkee, Roorkee 247667, Uttarakhand, India}

\date{\today}
          
\begin{abstract}
A systematic investigation of the odd-mass neutron-rich bromine isotopes $^{87-91}$Br has been carried out within the configuration-constrained cranked shell model (CSM) framework. The calculated kinematic moments of inertia and angular-momentum alignments reproduce the experimental trends with excellent agreement for the proposed quasiparticle configurations, thereby supporting the assigned band structures. The total Routhian surface calculations reveal pronounced $\gamma$-softness in $^{87}$Br, shape coexistence in $^{89-93}$Br, and a gradual prolate-to-oblate transition approaching $N=56$. These results demonstrate that the CSM provides a reliable description of rotational behavior and shape evolution in this mass region, achieving a better level of accuracy,  while offering a transparent interpretation of quasiparticle configurations and shape-driving effects.
\end{abstract}

\maketitle

\section{\label{sec:Intro}Introduction}
The shape evolution of a nucleus under increasing rotational frequency is a fascinating aspect of nuclear structure physics. At low rotational frequencies, nuclei often exhibit axial symmetry, with their deformation characterized by prolate or oblate shapes~\cite{bohr1998}. However, as the rotational frequency increases, the centrifugal force induced by rotation can drive the system toward more complex shapes, including triaxial deformation~\cite{Frauendorf2001}. This transition highlights the interaction between collective rotational motion and the intrinsic nuclear structure, determined by the distribution of nucleons in different orbitals. Understanding this evolution provides valuable insights into nuclear dynamics, shape coexistence, and the role of angular momentum in determining nuclear configurations. 

Investigating single-particle energies and collective behaviors in exotic, neutron-rich nuclei is a significant area of focus in nuclear structure research~\cite{sorlin2008nuclear,Paar_2007,nature2019}. The configuration-constrained cranked shell model (CSM) has proven to be an effective framework for understanding rotational bands, especially in systems where quasiparticle alignments and level crossings play a prominent role~\cite{bengtsson1979,NAZAREWICZ1985397,WYSS1988211,NAZAREWICZ1989285,SATULA199445,CC-CSM_Liang}. Medium-spin rotational states provide a unique perspective on the interplay between collective and single-particle dynamics, bridging the gap between low-spin states, dominated by shell effects, and high-spin states, where collective rotational behavior is more pronounced. In particular, odd-even nuclei, such as the neutron-rich Br isotopes, \(^{87, 89, 91, 93}\text{Br}\), offer an opportunity to explore complex nuclear phenomena due to the coupling between an unpaired valence proton and the core. These isotopes are characterized by relatively small quadrupole deformations (\(\beta_2 \sim 0.15\)), which enhance the sensitivity of their rotational behavior to specific single-particle configurations and level crossings.

Recent studies have extensively explored the high-spin and medium-spin states of neutron-deficient and neutron-rich nuclei, focusing on their rotational structures and collective behavior. In \( ^{75}\text{Br} \), high-spin structures were investigated through complementary models, uncovering shape coexistence and rigid rotational bands, with cranked-shell and particle-rotor coupling models aiding the interpretation~\cite{Br75paper}. Similarly, \( ^{73}\text{Br} \) exhibited octupole correlations and band termination at high spins, supported by Doppler-shift attenuation lifetime analysis and total Routhian surface (TRS) calculations~\cite{Br73paper}. Medium-spin and yrast structures in \( ^{87}\text{Br} \) and \( ^{89}\text{Br} \), populated via cold-neutron-induced fission, revealed enhanced collectivity and deviations from shell model predictions, attributed to the polarization effects of \( \pi g_{9/2} \) protons~\cite{Br87paper, Br89paper}. Comparative studies across \( ^{87{-}93}\text{Br} \) isotopes using large scale shell model (LSSM), and discrete nonorthogonal (DNO) shell model calculations identified a transition from prolate to oblate shapes, attributed to a balance of proton and neutron quadrupole deformations, signifying a pseudo-SU3 quadrupole regime~\cite{J.Dudouet}. 
Advanced approaches such as the Monte Carlo shell model and DNO-LSSM also reproduce these trends and reveal pseudo-SU(3) symmetry, moderate prolate deformation, and increasing $\gamma$-softness in the As isotopes~\cite{rezynkina2022structure}. Near the doubly magic $^{78}$Ni, studies of Zn, Ge, and Se isotopes show that while $N=50$ remains a robust shell closure in $^{80}$Zn, neutron addition induces deformation, triaxial softness, and shape coexistence, especially for $N=52$--54~\cite{taniuchi201978ni,shand2017shell}. Medium- and high-spin studies on Se, Sr, and Zr isotopes demonstrate a transition from near-spherical configurations at $N=50$ to pronounced deformation approaching $N=60$, largely attributed to excitations into the $g_{9/2}$ and intruder orbitals. In heavier regions, extensive spectroscopy of Cs isotopes, particularly $^{118}$Cs and $^{119}$Cs, has uncovered rich rotational structures, while lighter isotopes ($^{112\text{--}117}$Cs) remain less explored but show emerging collectivity with new bands recently established in $^{117}$Cs~\cite{jodidar2024collectivity}. Collectively, these studies emphasize the evolving shell structure, deformation patterns, and quasiparticle configurations that characterize nuclei away from closed shells in this mass region.

This work systematically investigates the medium-spin rotational bands in \(^{87, 89, 91}\text{Br}\). The spin and configuration assignments of these bands are confirmed by comparing the experimental moments of inertia and angular momentum alignments with theoretical predictions. TRS calculations trace the shape evolution of the nuclei, revealing a transition from prolate to oblate shapes, particularly when $N = 56$. This study aims at a detailed understanding of the rotational dynamics and shape evolution in neutron-rich Br isotopes, with potential implications for new configurations and phenomena in this mass region. The structure of this paper is organized to provide a clear progression of the framework, results, and conclusions. Section~\ref{sec:Theory} focuses on the theoretical framework that underlies this study. Here, we outline the calculation to describe nuclear spin states, including the configuration-constrained cranked shell model and the Strutinsky shell-correction method. Additionally, the methodology for calculating angular momentum alignments and the kinematic moment of inertia is explained. Section~\ref{sec:Disc} presents the detailed results of our study. It includes the deformation parameters chosen for all the odd isotopes of Br, along with the TRS calculated at four different rotational frequencies. These results are presented to gather valuable insights into the evolution of nuclear shapes and configurations under varying rotational conditions. The section also discusses the assignment of band configurations, which is crucial for understanding the observed nuclear spectra and their connection to the underlying quasiparticle structure. Finally, Section~\ref{sec:conclu} highlights the significant outcomes of this work in advancing our understanding of nuclear structure, particularly the interplay between deformation, rotational motion, and quasiparticle excitations.

\section{\label{sec:Theory}Theoretical Framework}
The description of nuclear spin states requires a framework that consistently incorporates both collective rotation and single-particle structure. The present theoretical approach is organized into three parts: (A) Configuration-constrained cranked shell model, (B) Strutinsky shell correction method, and (C) Angular-momentum alignment and kinematic moments of inertia (KMoI).
\subsection{\label{subsec:A}Configuration constrained cranked shell model}
The configuration-constrained cranked shell model provides such an approach by combining the cranking method with the Strutinsky shell-correction procedure~\cite{STRUTINSKY1967, STRUTINSKY1968}. The Strutinsky method forms the basis of the macroscopic--microscopic description of nuclear structure, in which the total energy of a nucleus is written as
\begin{equation}
    E_{\text{TOT}} = E_{\text{RLDM}} + \sum_{Z,N}\delta E ,
    \label{etot}
\end{equation}
where $E_{\text{RLDM}}$ represents the rotating liquid-drop energy, and $\delta E$ is the shell correction at finite angular momentum including pairing correlations, evaluated separately for protons ($Z$) and neutrons ($N$). The rotating liquid-drop energy of a triaxially deformed nucleus is given by~\cite{NEERG,BENGTSSON1975301,aru2004,aru2005}
\begin{equation}
    E_{\text{RLDM}} = E_{\text{LDM}} + \frac{1}{2}\omega^2 \mathscr{I}_{\text{rig}}(\beta,\gamma) ,
    \label{erldm}
\end{equation}
where $E_{\text{LDM}}$ is the liquid-drop energy, $\mathscr{I}_{\text{rig}}(\beta,\gamma)$ is the rigid-body moment of inertia with respect to the rotation axis, and $\omega$ is the rotational frequency. The nuclear shape is described by the deformation parameters $\beta = (\beta_2,\beta_4)$ and $\gamma$, where $\beta_2$ is the quadrupole deformation, $\beta_4$ denotes the hexadecapole correction to the shape, and $\gamma$ specifies the degree of triaxiality.  

The $E_{\text{LDM}}$ is evaluated by considering only the surface and Coulomb contributions~\cite{G_Shanmugam2001,Shanmugam1999}, as the volume, asymmetry, and pairing terms do not vary with deformation. For a triaxially deformed shape, $E_{\text{LDM}}$ is expressed as
\begin{equation}
\begin{split}
    E_{\text{LDM}}(\beta,\gamma) & = E_{\text{Surf}}(\beta,\gamma) + E_{\text{Coul}}(\beta,\gamma)    \\ & = \left\{[B_s(\beta,\gamma) - 1] + 2\chi \,[B_c(\beta,\gamma) - 1]\right\} a_s A^{2/3},
    \label{eldm}
\end{split}
\end{equation}
where $B_s(\beta,\gamma)$ and $B_c(\beta,\gamma)$ are dimensionless shape-dependent functions. The factor $B_s(\beta,\gamma)$ represents the ratio of the surface area of the deformed shape to that of the area of the sphere of the same volume, while $B_c(\beta,\gamma)$ denotes the ratio of the Coulomb energy of the deformed charge distribution to that of the spherical shape. The fissility parameter $\chi$ and surface energy coefficient $a_s$ are chosen to be $\frac{Z^2}{45A}$~\cite{funnyhills} and $19.7$ MeV~\cite{Shanmugam1999}, where $A$ is the mass number.

The evaluation of the correction term $\delta E$ in Eq.~(\ref{etot}) obtained through the Strutinsky shell-correction method (detailed in Sec.~\ref{subsec:B}) requires accurate knowledge of the underlying single-particle spectrum of the nucleus. For this purpose, we construct the single-particle Hamiltonian from a realistic mean-field potential, which in the present work is the triaxial Woods-Saxon (WS) potential with universal parametrization~\cite{universal_WS}. This potential is expressed as a sum of kinetic energy operator $T$, central mean-field potential $V_{\text{WS}}$, the Coulomb interaction term $V_{\text{Coul}}$, and the spin-orbit potential $V_{\text{S.O.}}$, given by~\cite{swati_2017, Siwach_2020,sharma2023giant}
\begin{equation}
  H_\text{WS} = T +  V_{\text{WS}}(\vec{r},\beta,\gamma) + V_{\text{Coul}}(\vec{r},\beta,\gamma) + V_{\text{S.O.}}(\vec{r},\beta,\gamma).
  \label{hws}
\end{equation}

The cranked shell model Hamiltonian of a triaxially deformed nucleus in the rotating frame is written as~\cite{bengtsson1979, aru2005, Zhang_2013}
\begin{equation}
H_{\text{CSM}} = H_{\text{WS}} + H_{\text{P}} - \omega J_x,
\label{hcsm}
\end{equation}
where $H_{\text{WS}}$ is the single-particle Wood-Saxon Hamiltonian given in Eq.~(\ref{hws}) and $H_\text{P}$ is the pairing Hamiltonian. The term $-\omega J_x$ corresponds to the cranking term, with $\omega$ denoting the rotational frequency and $J_x$ is the angular momentum operator about the $x$-axis. 

For a fixed deformation \((\beta,\gamma)\) and rotational frequency \(\omega\) the computation proceeds in two stages:

(i) \emph{Single-particle Routhians:} The cranked single-particle Hamiltonian $h'(\omega;\beta,\gamma) \equiv H_{\mathrm{WS}} - \omega J_x$ is diagonalized in a suitably large deformed harmonic-oscillator basis to obtain the cranked single-particle Routhians \(e^{'}_i(\omega;\beta,\gamma)\) and their eigenfunctions. These Routhians include the Coriolis shifts and provide the level scheme that enters both the shell-correction and the pairing calculations. Practical implementations follow established prescriptions for basis truncation and Woods–Saxon parametrization~\cite{BENGTSSONRAGNARSSON1985}.  

(ii) \emph{Quasiparticle Routhians:} The quasiparticle Routhians \(E^{'}_\alpha(\omega;\beta,\gamma)\) are obtained by incorporating pairing correlations into the cranked single-particle Hamiltonian as given in Eq.~(\ref{hcsm}). Pairing correlations at each rotational frequency and deformation point $(\omega;\,\beta,\gamma)$ are treated by solving the BCS equations~\cite{bcs_superconductivity}. Using the cranked single-particle spectrum as input, the BCS procedure yields the pairing gaps, chemical potentials, and quasiparticle occupation amplitudes~\cite{bengtsson1979, NAZAREWICZ1985397}. The quasiparticle Routhians are then obtained from these solutions and used in the evaluation of Strutinsky shell corrections, the construction of blocked multi-quasiparticle configurations, and the analysis of alignments and band-crossing phenomena.

In the configuration-constrained cranked shell model~\cite{bengtsson1979,NAZAREWICZ1989285,WYSS1988211,NAZAREWICZ1985397}, a specific rotational band is generated by fixing the occupancy of selected quasiparticle orbitals that characterize the underlying intrinsic configuration. The relevant quasiparticle orbitals are blocked, and the pairing problem is re-solved self-consistently for all remaining particles. For each blocked configuration, one evaluates the total Routhian $E_{\text{TOT}}$ as a function of \((\beta,\gamma)\) and minimizes to obtain the configuration-constrained equilibrium shape (TRS) at fixed \(\omega\). The yrast configuration at a given \(\omega\) (or \(I\)) is the configuration with the lowest total Routhian.

\subsection{\label{subsec:B}Strutinsky shell correction method}

The microscopic correction $\delta E$ entering Eq.~(\ref{etot}) is evaluated as a sum of a shell (single-particle) correction $\delta E_{\mathrm{shell}}$ and a pairing correction $\delta E_{\mathrm{pair}}$. For each rotational frequency and deformation point $(\omega;\,\beta,\gamma)$ it is given by
\begin{equation}
  \delta E \;=\; \delta E_{\mathrm{shell}} \;+\; \delta E_{\mathrm{pair}}.
  \label{eq:6}
\end{equation}

\paragraph{Strutinsky shell correction:}
The shell correction at finite rotation is obtained by subtracting the Strutinsky-smoothed single-particle energy $\widetilde{E}_{sp}$, incorporating the averaged spectrum, curvature term, and smoothed rotational contribution, from the microscopic energy $E_{sp}$ built from the occupied cranked single-particle Routhians and rotational term~\cite{BENGTSSON1975301,NAZAREWICZ1989285,aru2004,aru2005}.
\begin{equation}
  \delta E_{\mathrm{shell}}
  \;=\; E_{sp} \;-\; \widetilde{E}_{sp},
  \label{eq:7}
\end{equation}
\begin{equation}
  E_{sp}\;=\; \sum_{i}^{\mathrm{occ}} e'_i n_i \;+\; \omega I_{sp},
  \label{eq:8}
\end{equation}
\begin{equation}
   \widetilde{E}_{sp}\;=\; \sum_{i}^{\infty} e'_i \widetilde{n}_i - \gamma_s\widetilde{M}_c \;+\; \omega \widetilde{I},
  \label{eq:9}
\end{equation}
here $\gamma_s$ is the smearing parameter, $n_i$ is the occupation numbers which follow the Fermi-Dirac distribution, and $\tilde{n}_i$ is the smoothed occupation number given by
\begin{equation}
\widetilde{n}_i = \frac{1}{2} \Big[ 1 + \mathrm{erf}(x_i) \Big] - \frac{1}{\sqrt{\pi}} \exp(-x_i^2) \sum_{m=1}^{p} C_m H_{m-1}(x_i),
\label{eq:10}
\end{equation}
where $x_i = (e' - e'_i)/\gamma_s$, $\mathrm{erf}$ is the error function, $H_m$ denotes the Hermite polynomial of order $m$, and the coefficients $C_m=\left\{
\begin{array}{cl}
\frac{(-1)^{m/2}}{2^m(m/2)!}\  & \mathrm{if}\ m\ \text{is  even} \\
0 & \mathrm{if}\ m\ \text{is odd}
\end{array} 
\right. ,\ $ and the parameter $p$ sets both the order of the polynomial and the degree of smoothing.\\
The Strutinsky curvature correction is expressed as~\cite{greiner1996nuclear}
\begin{equation}
\widetilde{M}_c = \frac{1}{2\sqrt{\pi}} \sum_{i=1}^{\infty} C_p H_p(\widetilde{x}) \exp(-\widetilde{x}^2),
\label{eq:11}
\end{equation}
where $\widetilde{x} = (\widetilde{\lambda} - e'_i)/\gamma_s$, and the smoothed Fermi energy $\widetilde{\lambda}$ is determined by enforcing the particle-number constraint $\widetilde{\mathcal{N}}(\widetilde{\lambda}) = N_p$, with $\widetilde{\mathcal{N}} = \sum_i^\infty \widetilde{n}_i$. The inclusion of the curvature correction reduces the sensitivity of the shell correction to the smearing width $\gamma_s$, thereby producing a broader plateau in energy.

The single-particle occupation numbers and Routhians depend on \(\omega\), the expectation value of the angular momentum (about the rotation axis) contains a fluctuating (shell) part as well when considering a fixed particle number. Accordingly, a shell correction can be introduced for the total spin, analogous to the shell correction in the energy~\cite{BRACK198135, aru2004, aru2005}.
\begin{align}
\label{eq:jx}
  I_{sp} &\;=\; \langle J_x \rangle \;=\; \sum_{i}^{\mathrm{occ}}  \langle j_{x,i} \rangle n_i , \\
  \widetilde{I} &\;=\; \sum_{i=1}^{\infty} \langle j_{x,i} \rangle \widetilde{n}_i,
\end{align}
where  $\langle j_{x,i} \rangle$ is the single-particle projection for the \(i\)-th Routhian.

\paragraph{Pairing correction:}
The correction in the pairing energy $\delta E_{pc}$ is given by~\cite{Pairing_ref1} 
\begin{equation}
\delta E_{pc} = E_{pc} - \widetilde{E}_{pc},
\label{eq:12}
\end{equation}
where,
\begin{equation}
    E_{pc} =  2 \!\left( \sum_{k=1}^{N_p} \epsilon_k v_{k}^{2} - \sum_{k=1}^{\frac{N_p}{2}} \epsilon_k \!\right) - \frac{\Delta^2}{G} -G \!\left(\sum_{k=1}^{N_p} v_{k}^{4} - \sum_{k=1}^{\frac{N_p}{2}} 1\!\right),
    \label{eq:13}
\end{equation}

\begin{equation}
\begin{aligned}
\widetilde{E}_{pc} ={}& 
-\frac{1}{2}\frac{N_p^2}{\widetilde{g}(\widetilde{\lambda})}
\left[\left(1+\frac{\widetilde{g}(\widetilde{\lambda})\,\widetilde{\Delta}}{N_p}\right)^{1/2}-1\right] \\
&\quad
+\,\widetilde{g}(\widetilde{\lambda})\,\widetilde{\Delta}G\,
\tan^{-1}\!\left(\frac{N_p}{\widetilde{g}(\widetilde{\lambda})\,\widetilde{\Delta}}\right),
\end{aligned}
\label{eq:14}
\end{equation}
where $\epsilon_k$ are the single-particle energies calculated using the triaxial Woods-Saxon potential. $v_k^2$, $\Delta$, and $\lambda$ are the BCS occupation probability, pairing gap, and chemical potential, respectively. $\widetilde{g}(\widetilde{\lambda})$ denotes the average level density at the Fermi level, which
is calculated using
\begin{equation}
\widetilde{g}(e) = \frac1{\gamma_s  \sqrt{\pi }}\sum_{i=1}^{\infty} \exp (-x_i^2)\sum_{m=0}^pC_mH_{m}(x_i),
\label{eq:15}
\end{equation}
$\widetilde{\Delta}$ is taken as $\frac{12}{\sqrt{A}}$ in our calculations. The pairing force strength ($G$) is chosen for protons and neutrons as~\cite{Pair_ref2}
\begin{equation}
\begin{aligned}
&G_p = [g_0^p + g_1^p(N - Z)]/A, \\
&G_n = [g_0^n - g_1^n(N - Z)]/A,
\end{aligned}
\end{equation}
where $g_0^p$, $g_1^p$, $g_0^n$, and $g_1^n$ are chosen as 17.90, 0.176, 18.95, and 0.078, respectively.

\subsection{Angular-momentum alignment and KMoI}
The kinematic moment of inertia (KMoI), $\mathcal{J}^{(1)}$, quantifies the rotational response of the nuclear system by relating the aligned angular momentum to the applied rotational frequency. Within the CSM, it is defined for a quasiparticle configuration as~\cite{bengtsson1979}  

\begin{equation}
\mathcal{J}^{(1)}(\omega) = \frac{\langle J_x \rangle}{\omega},
\label{eq:kmoi}
\end{equation}

Physically, \(\mathcal{J}^{(1)}(\omega)\) reflects the first-order buildup of angular momentum with increasing \(\omega\). Pairing correlations reduce \(\langle J_x \rangle\), lowering \(\mathcal{J}^{(1)}\), while quasiparticle blocking (e.g., in odd-A or multi-quasiparticle configurations) reduces pairing and increases \(\mathcal{J}^{(1)}\), producing odd–even staggering. Sharp increases (backbending) mark the alignment of high-\(j\) intruder quasiparticles crossing the Fermi surface~\cite{STEPHENS1972}.

Experimentally, \(\mathcal{J}^{(1)}\) and \(\hbar\omega\) are extracted from \(\gamma\)-ray transition energies $E_\gamma$
\begin{equation}
\frac{\mathcal{J}^{(1)}(I)}{\hbar^2} = \frac{2I + 1}{E_\gamma(I+1 \rightarrow I - 1)},
\label{eq:expMoI}
\end{equation}
\begin{equation}
\hbar\omega(I) = \frac{E_\gamma(I+1 \rightarrow I-1)}{I_x(I+1) - I_x(I-1)},
\label{eq:exp_w}
\end{equation}
where $I$ is the total angular momentum and \(I_x(I) = \sqrt{(I+1/2)^2 - K^2}\), $K$ is the projection of $I$ onto the symmetry axis. These equations allow direct comparison between the theory and experimental rotational spectra.

\section{\label{sec:Disc}Results and Discussions}

The configuration constrained cranked shell model calculations are carried out for $^{87,89,91}$Br isotopes and the resulting total Routhian surface (TRS), alignments, and kinetic moment of inertia are presented below. 

\subsection{Total Routhian surfaces}
\begin{figure*}[t]
    \centering
        \includegraphics[scale=0.32]{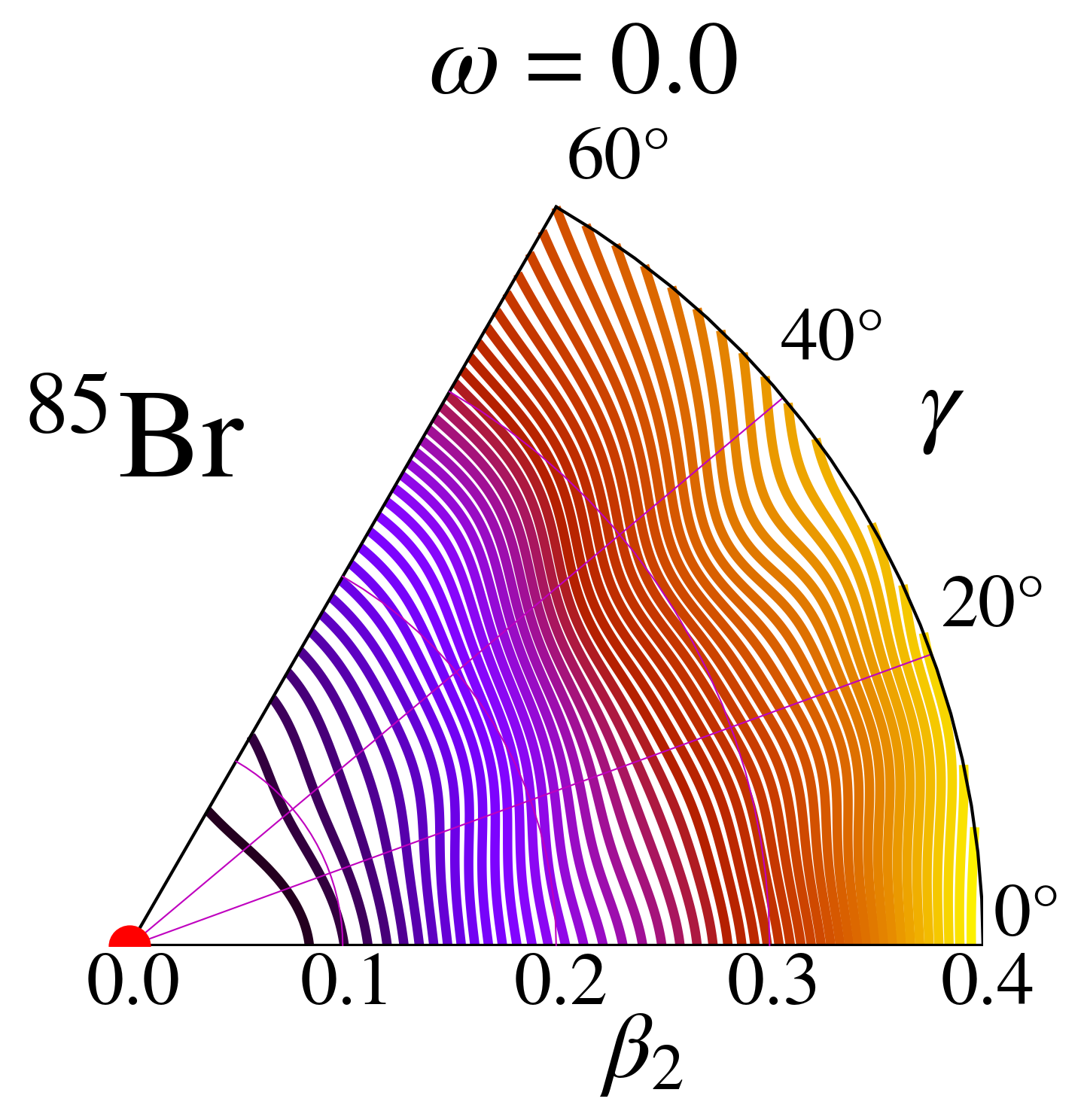}
        \includegraphics[scale=0.32]{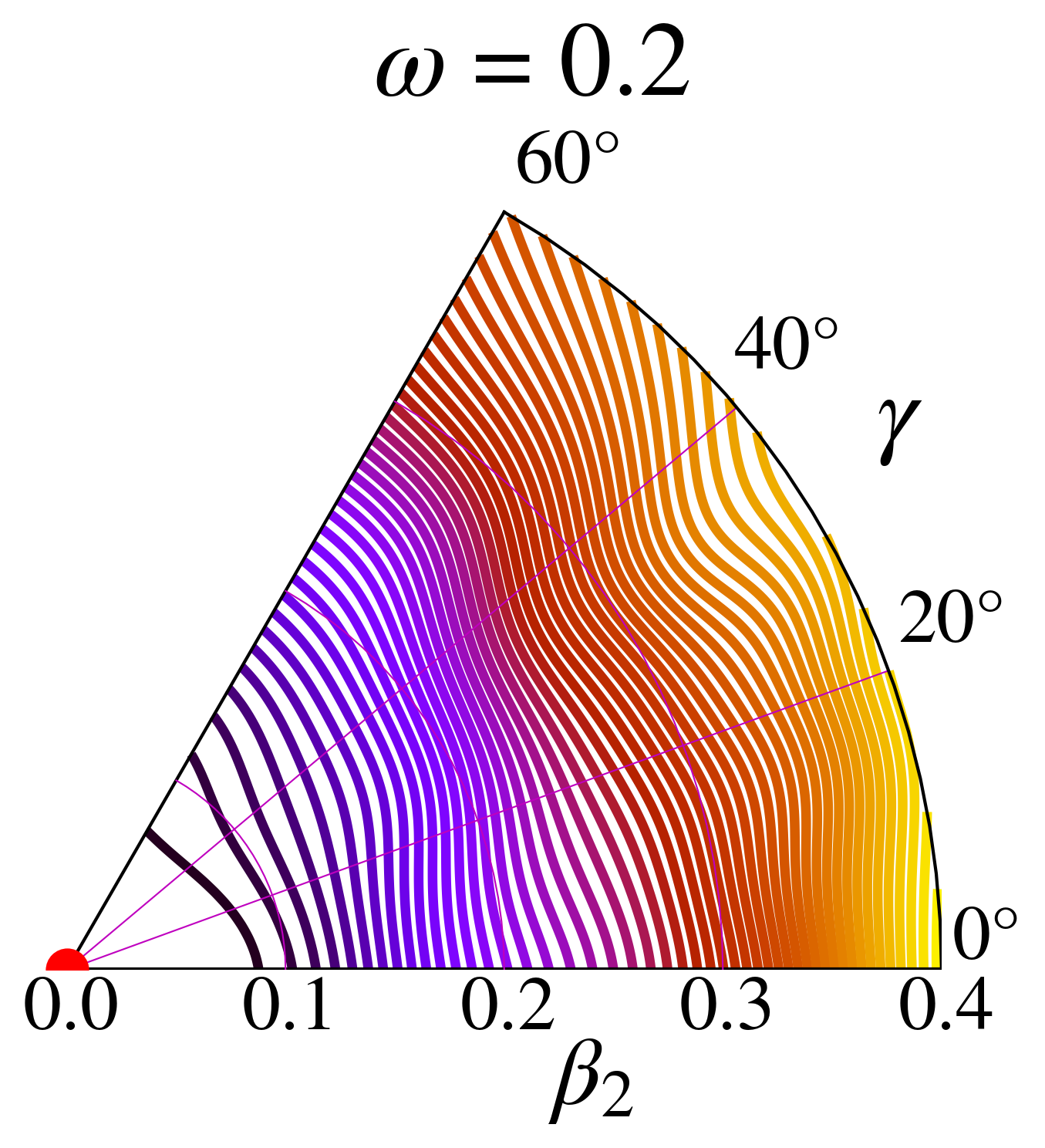}
        \includegraphics[scale=0.32]{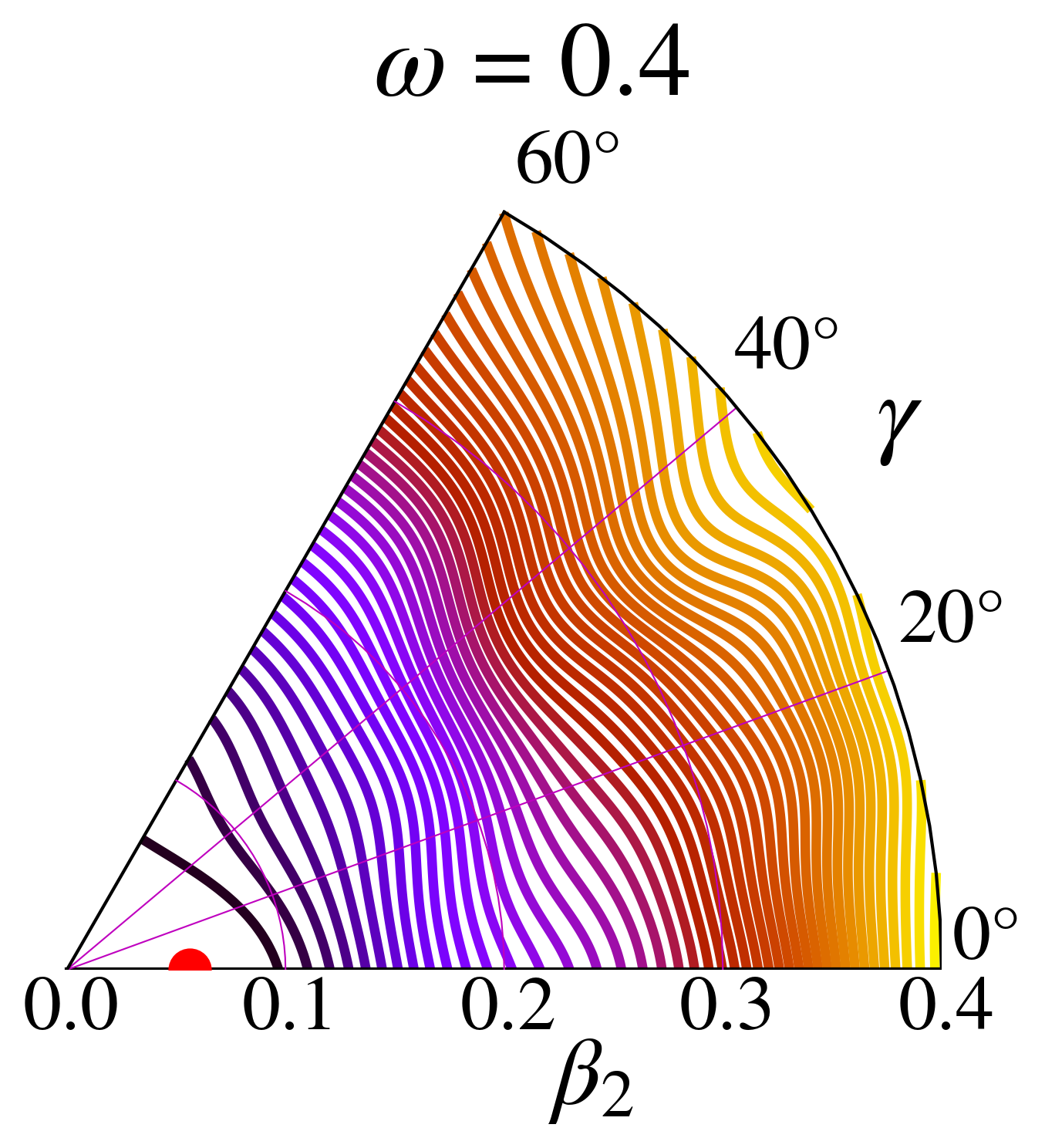}
        \includegraphics[scale=0.32]{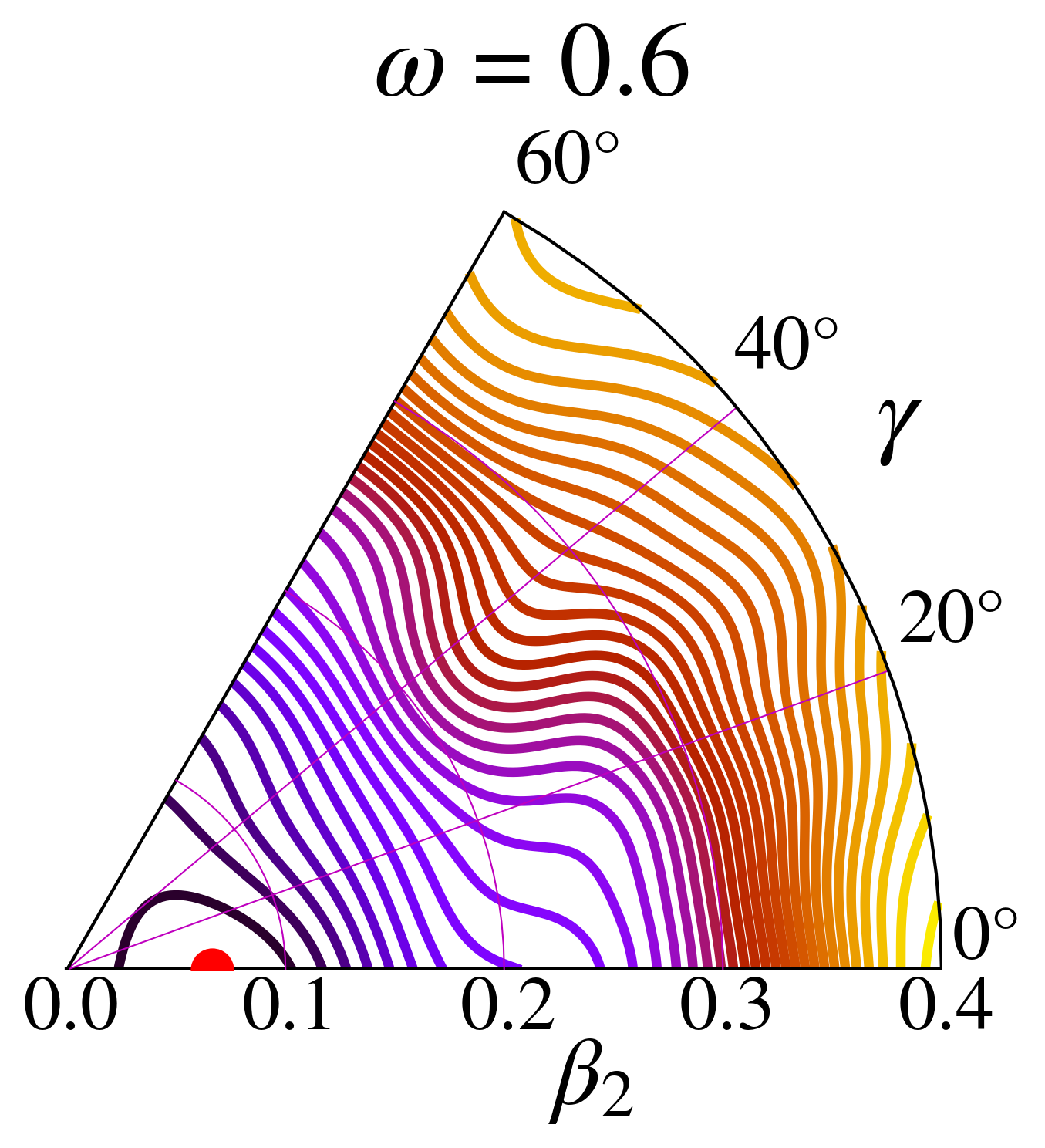}
        
        \includegraphics[scale=0.32]{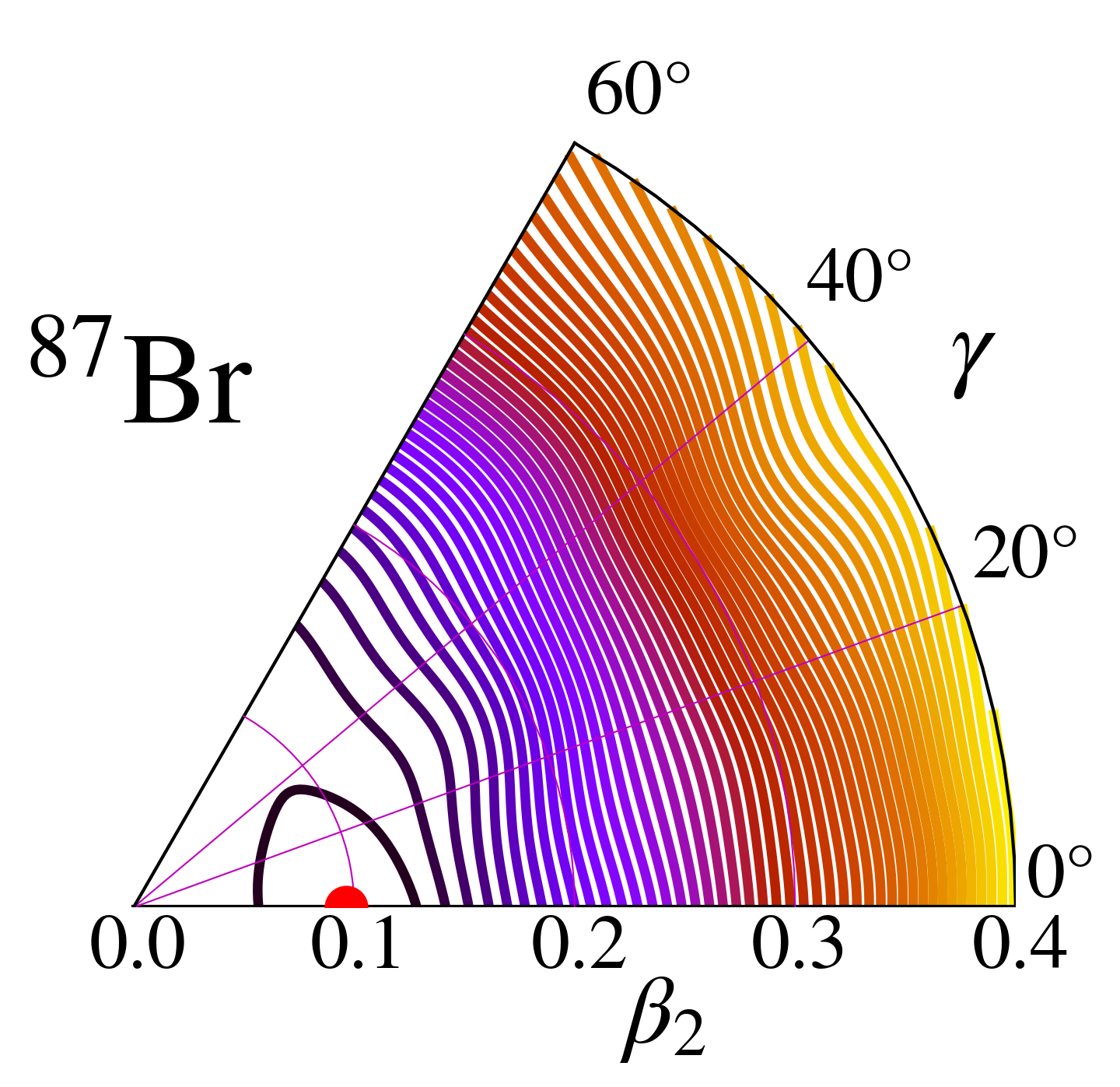}
        \includegraphics[scale=0.32]{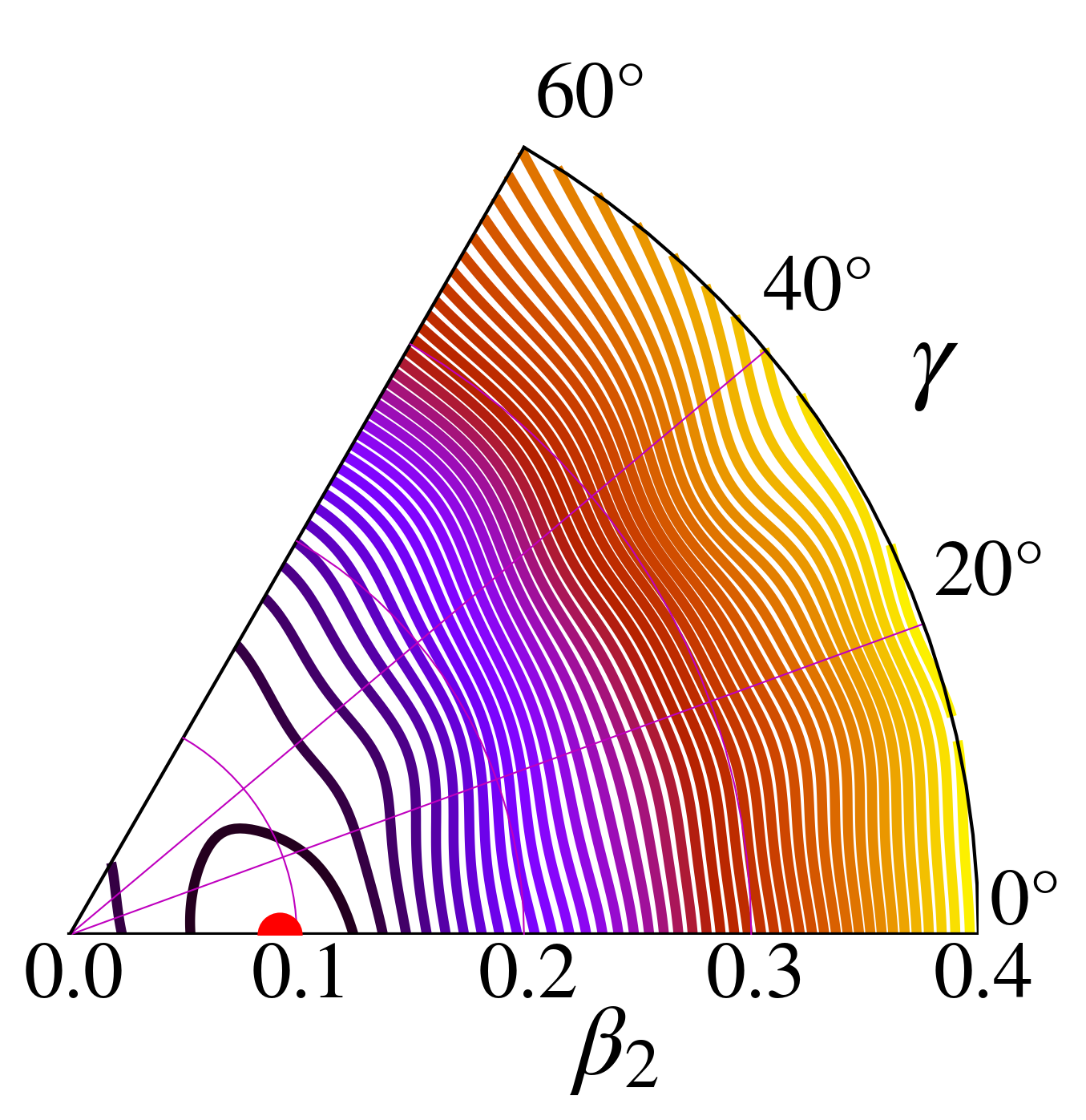}
        \includegraphics[scale=0.32]{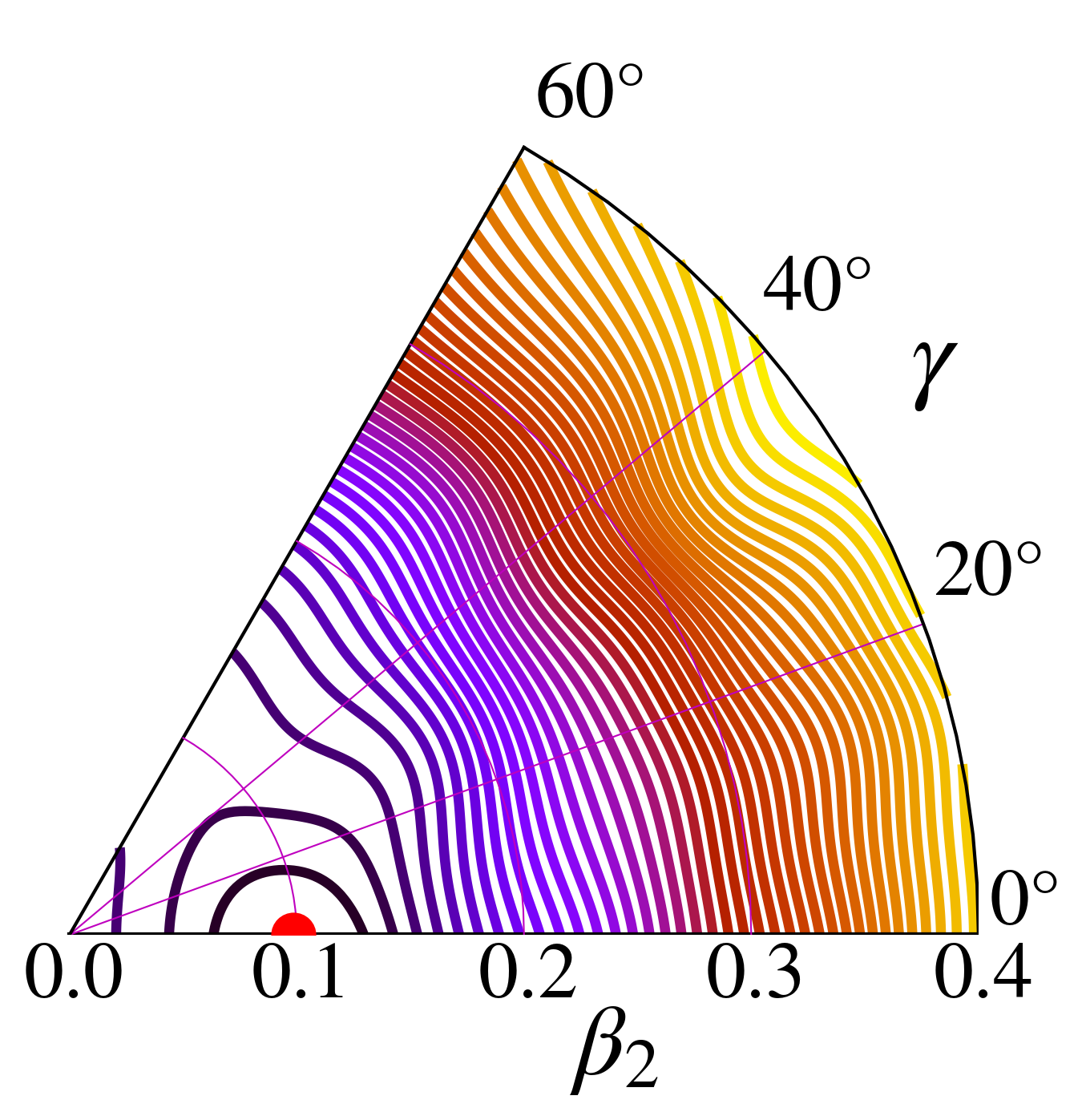}
        \includegraphics[scale=0.32]{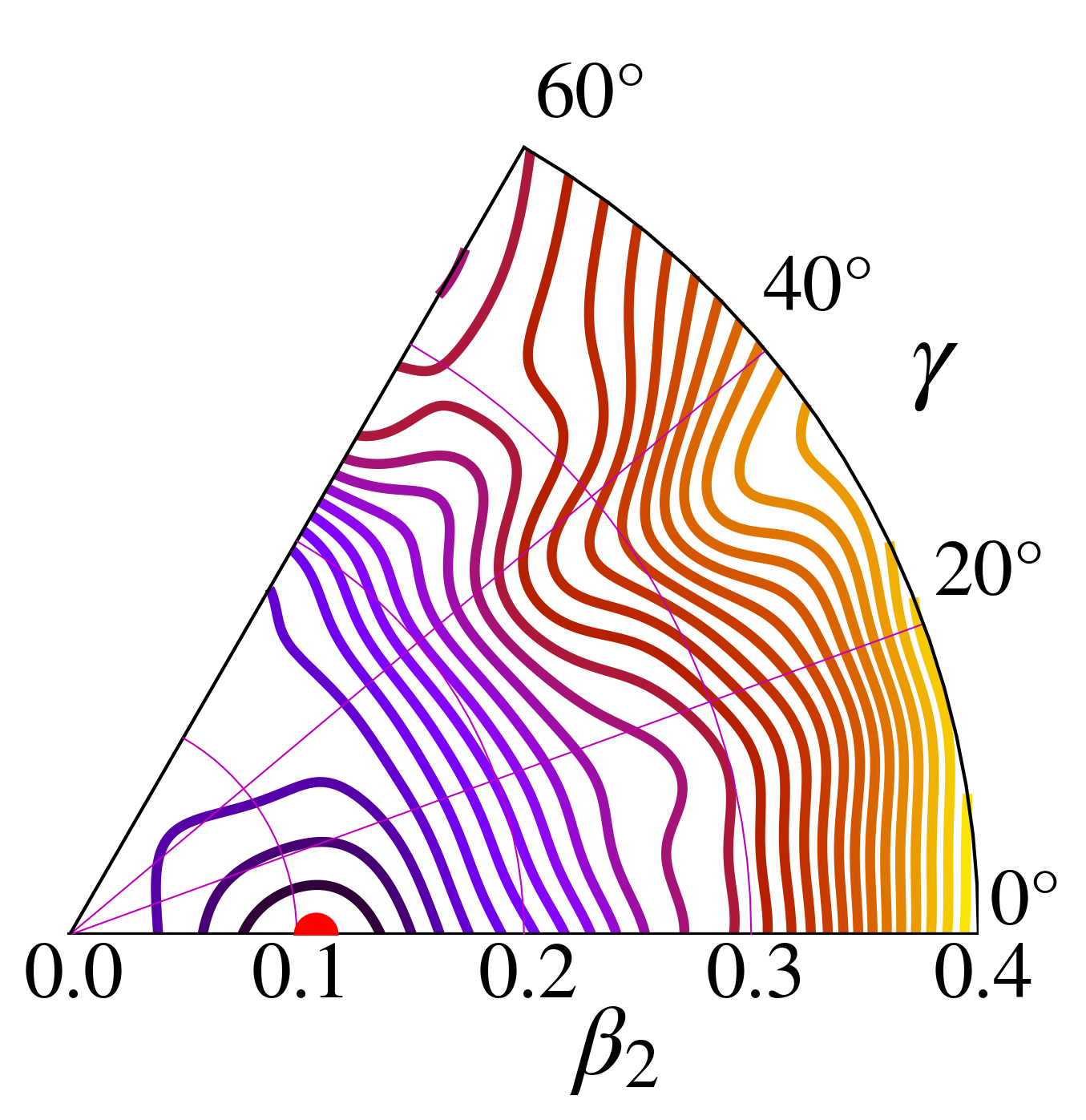}
        
        \includegraphics[scale=0.32]{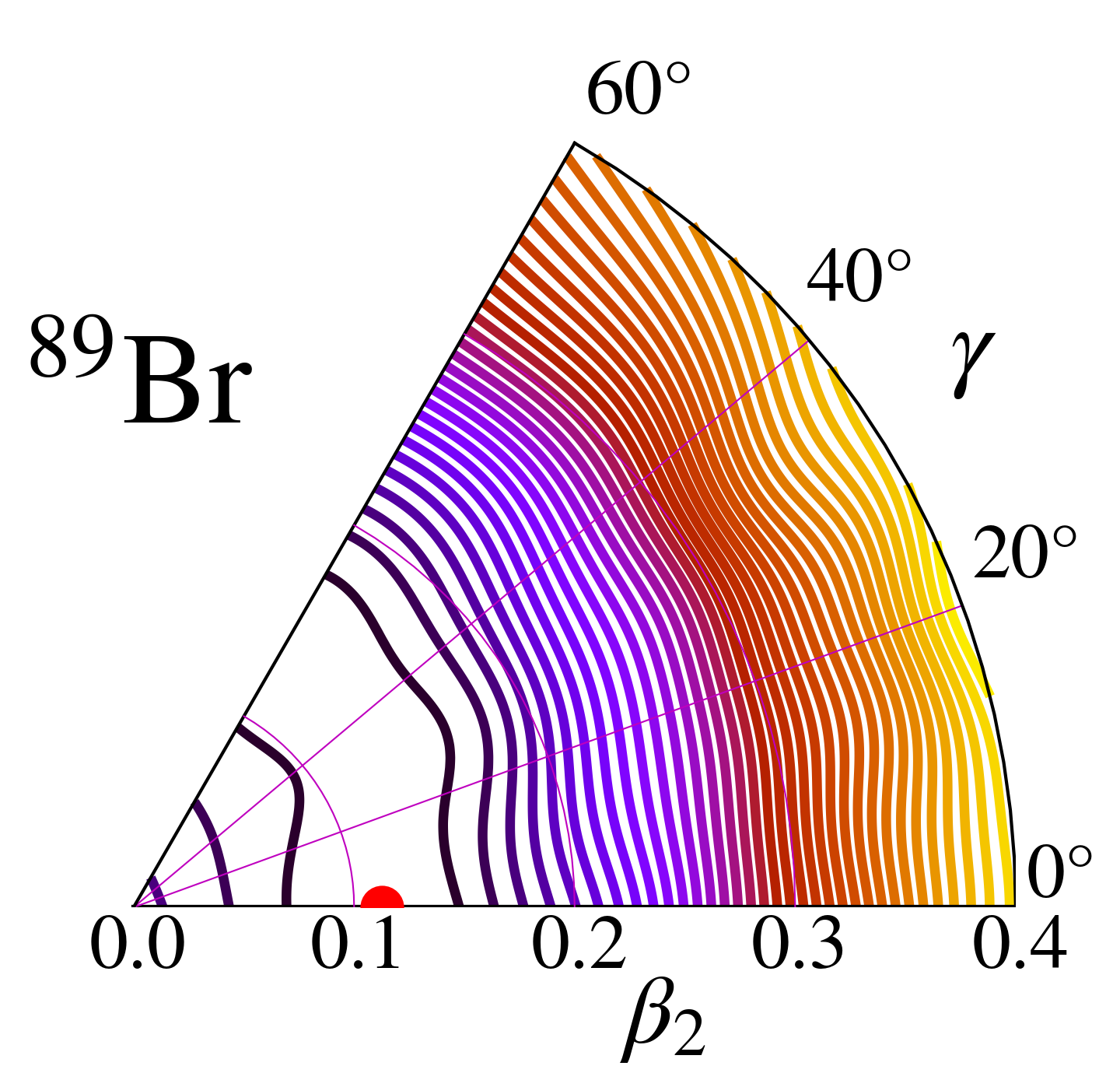}
        \includegraphics[scale=0.32]{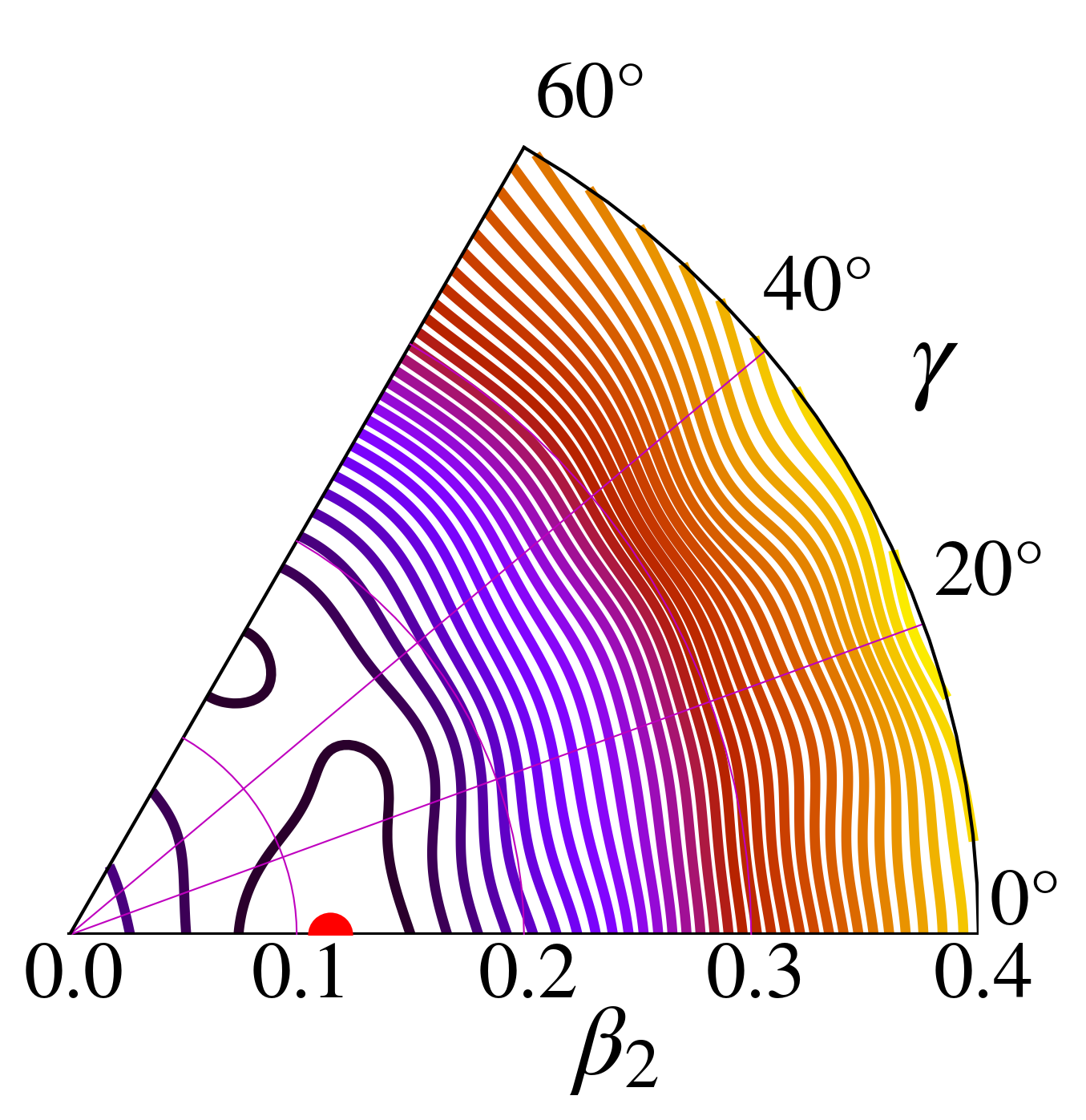}
        \includegraphics[scale=0.32]{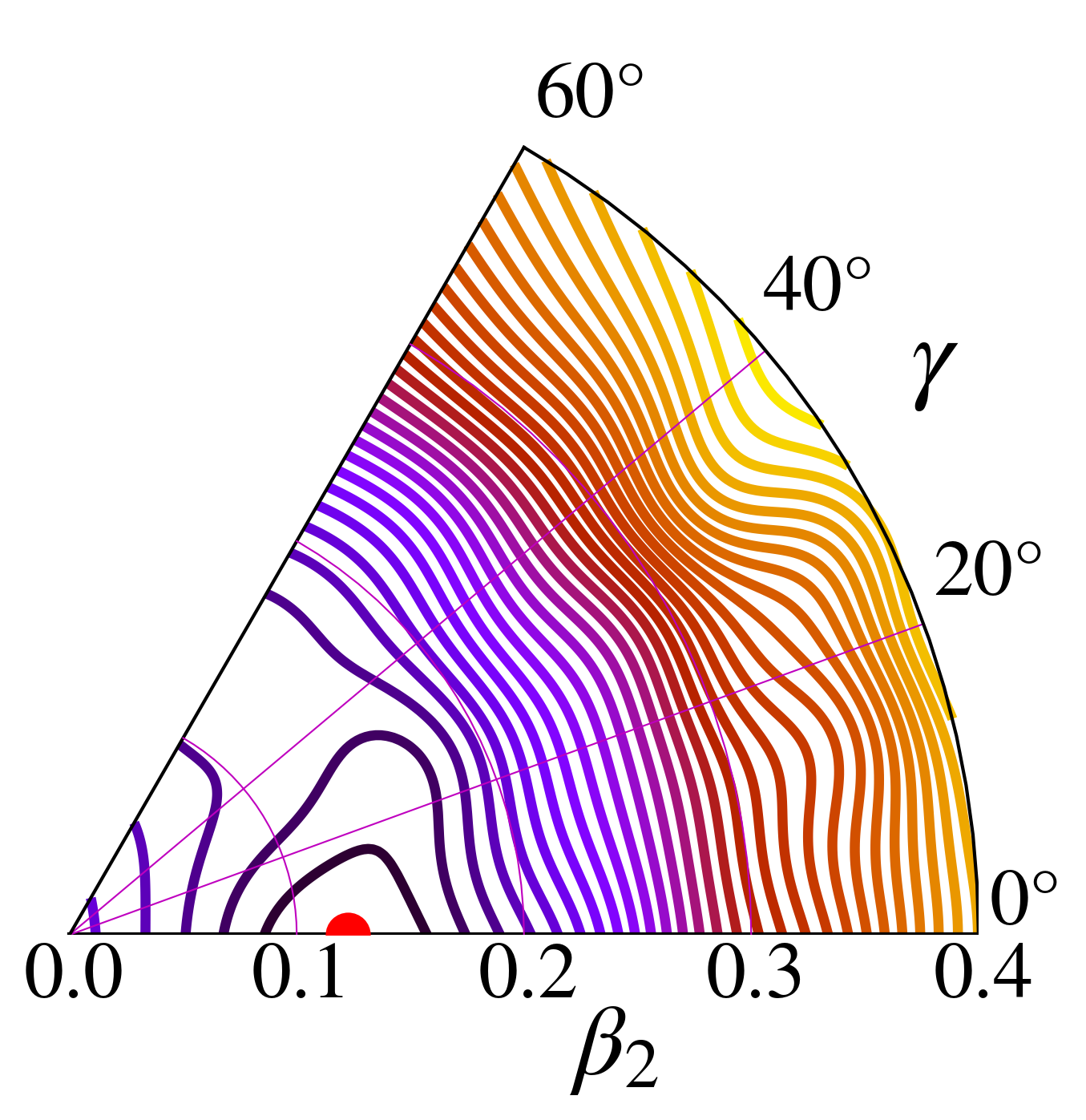}
        \includegraphics[scale=0.32]{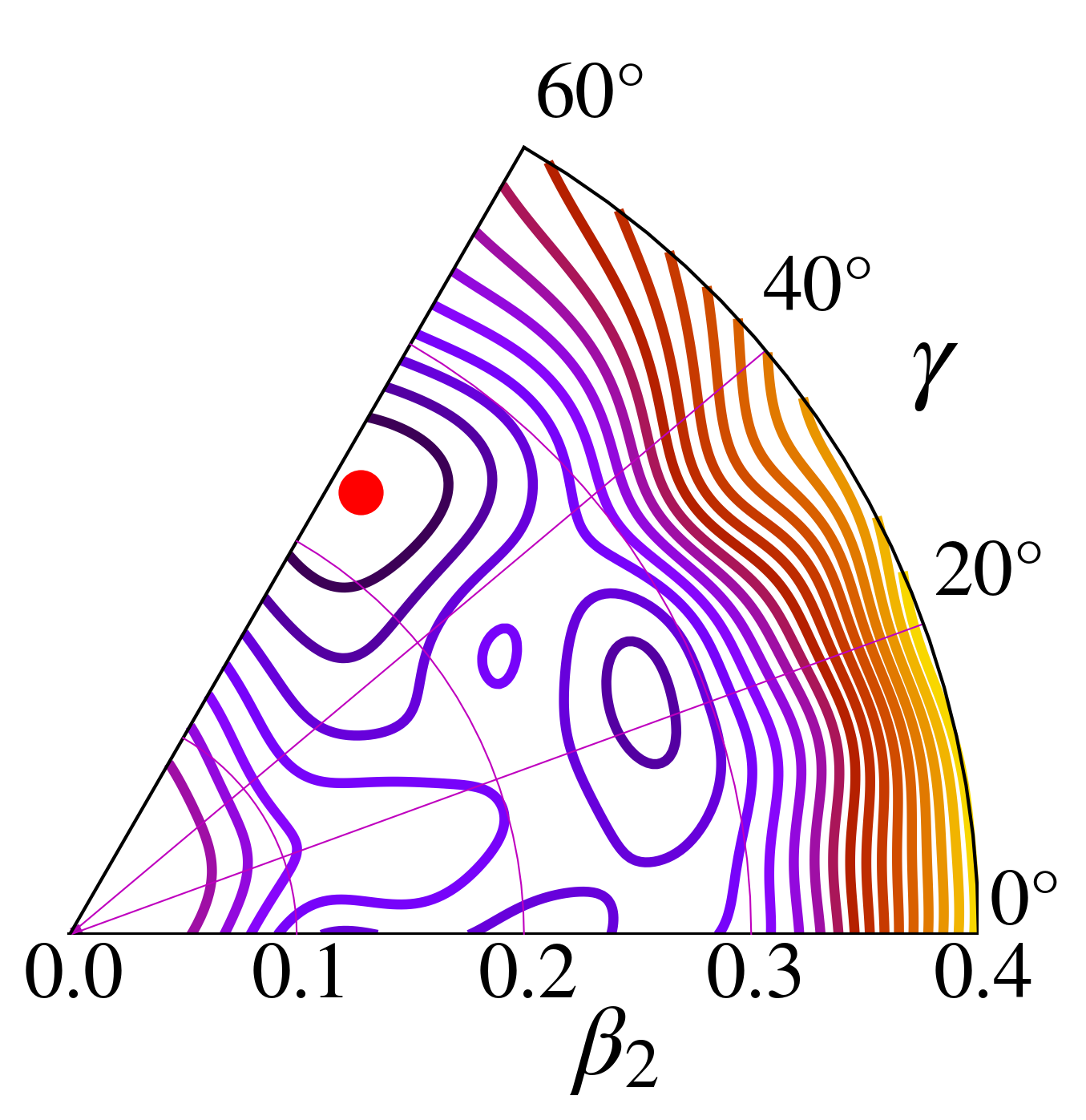}
        
        \includegraphics[scale=0.32]{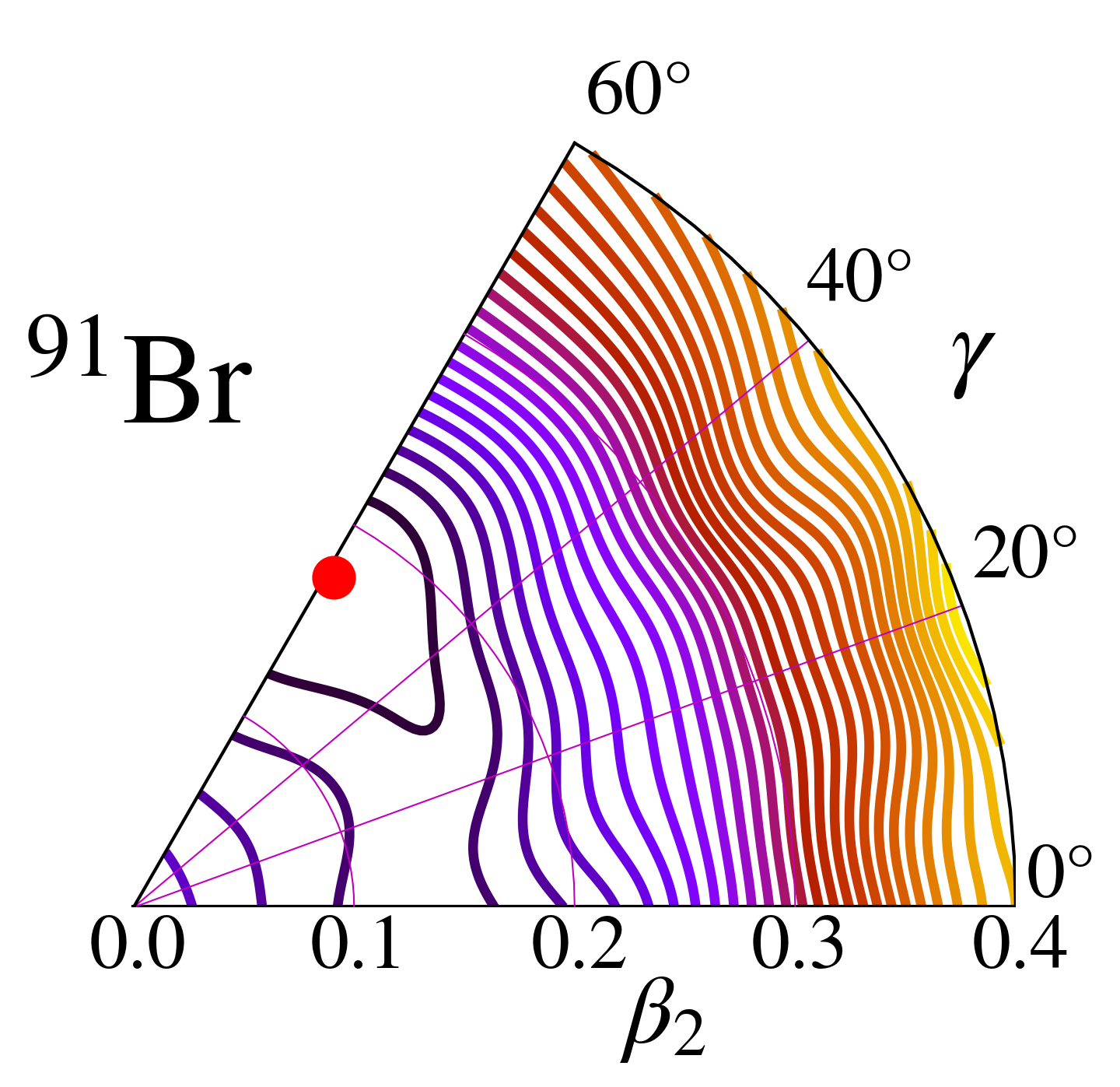}
        \includegraphics[scale=0.32]{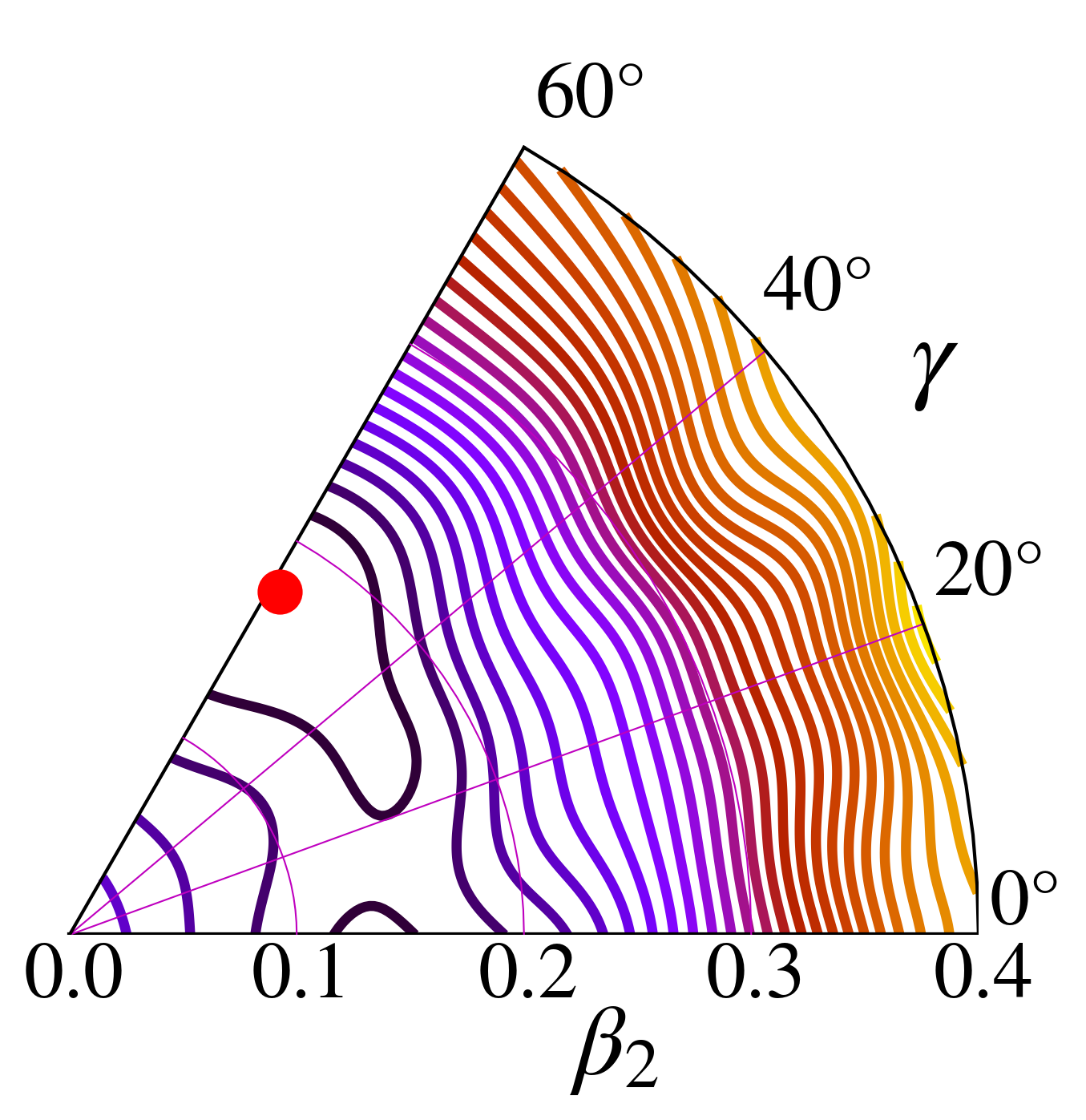}
        \includegraphics[scale=0.32]{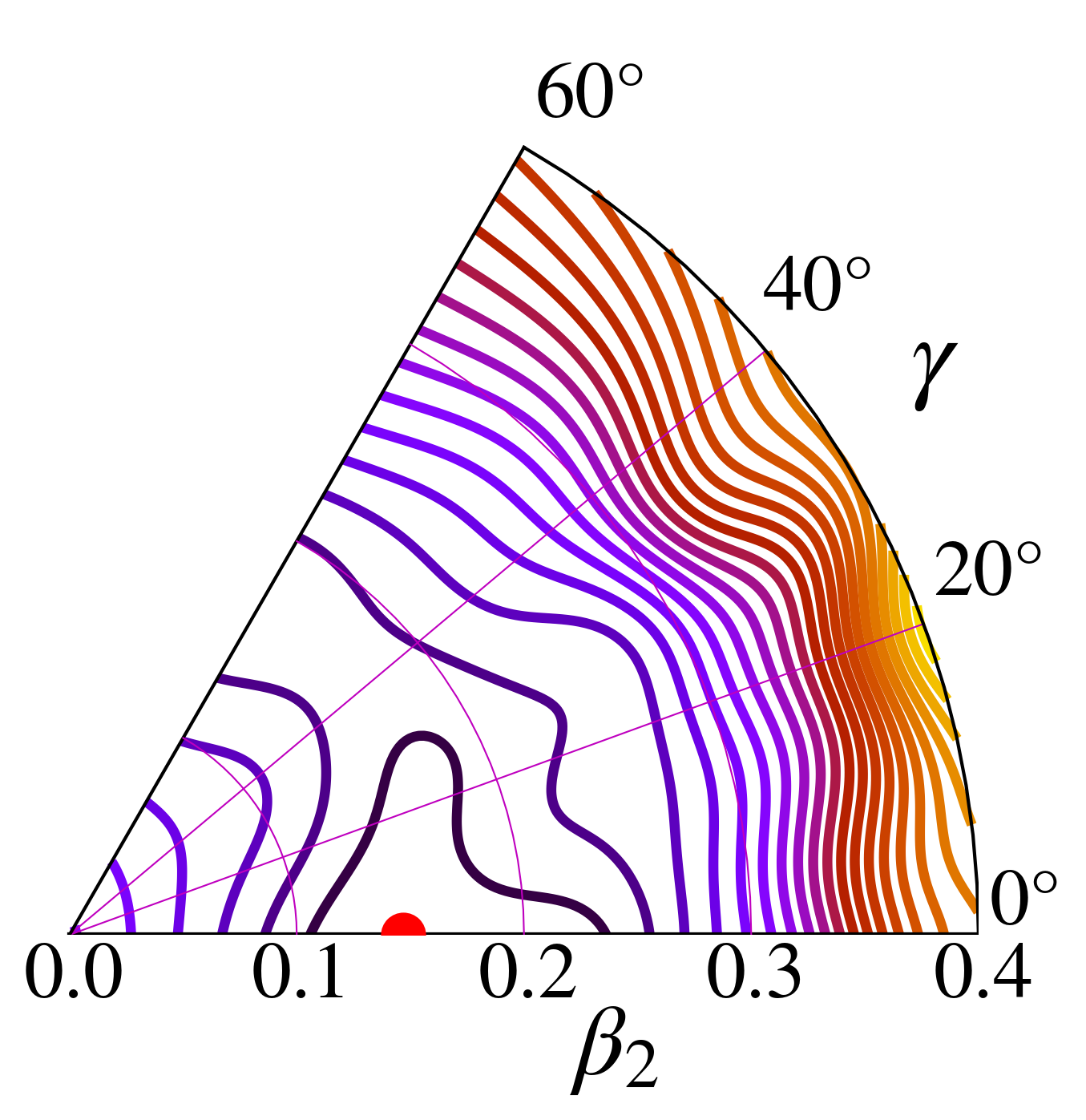}
        \includegraphics[scale=0.32]{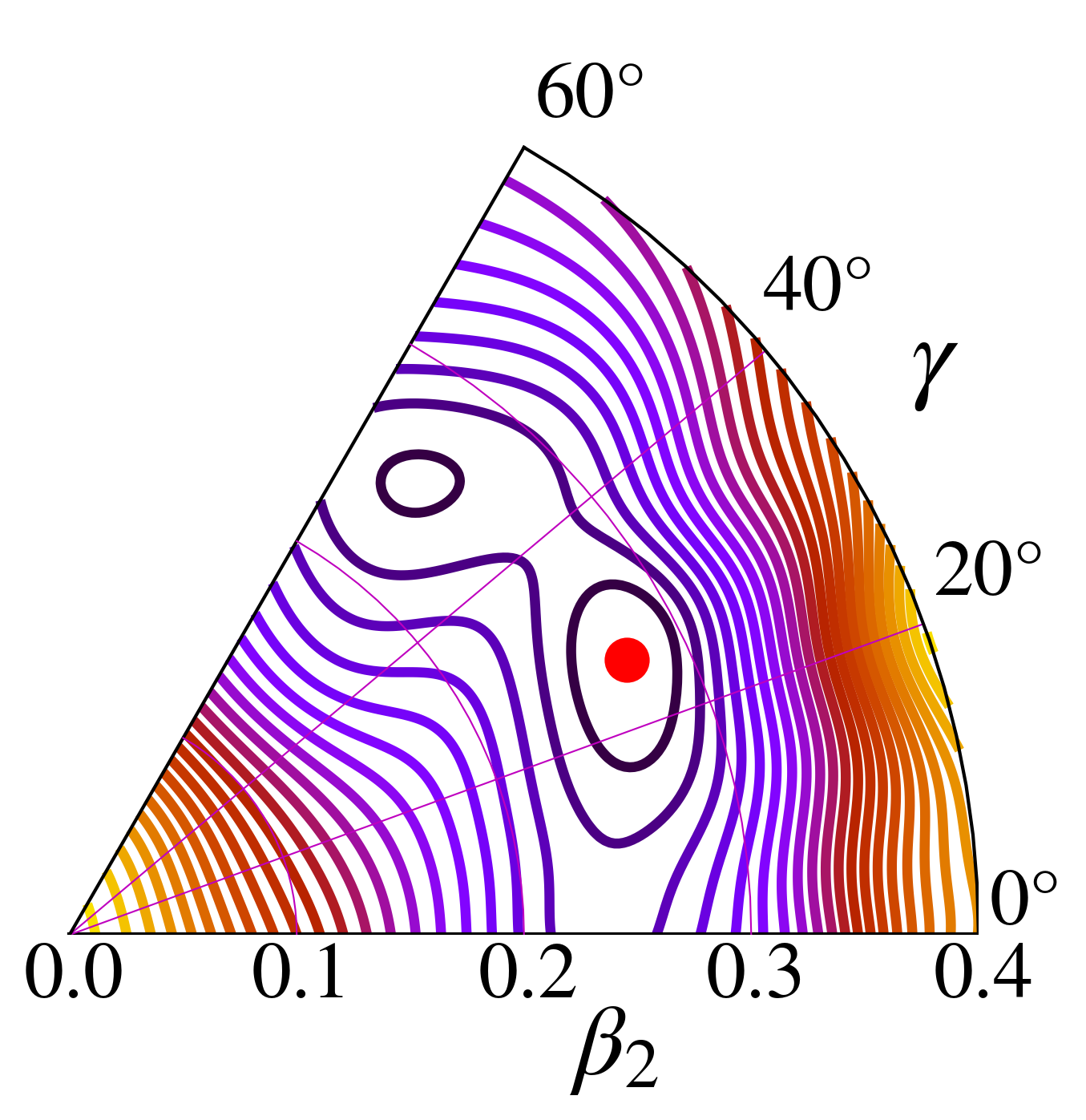}

        \includegraphics[scale=0.32]{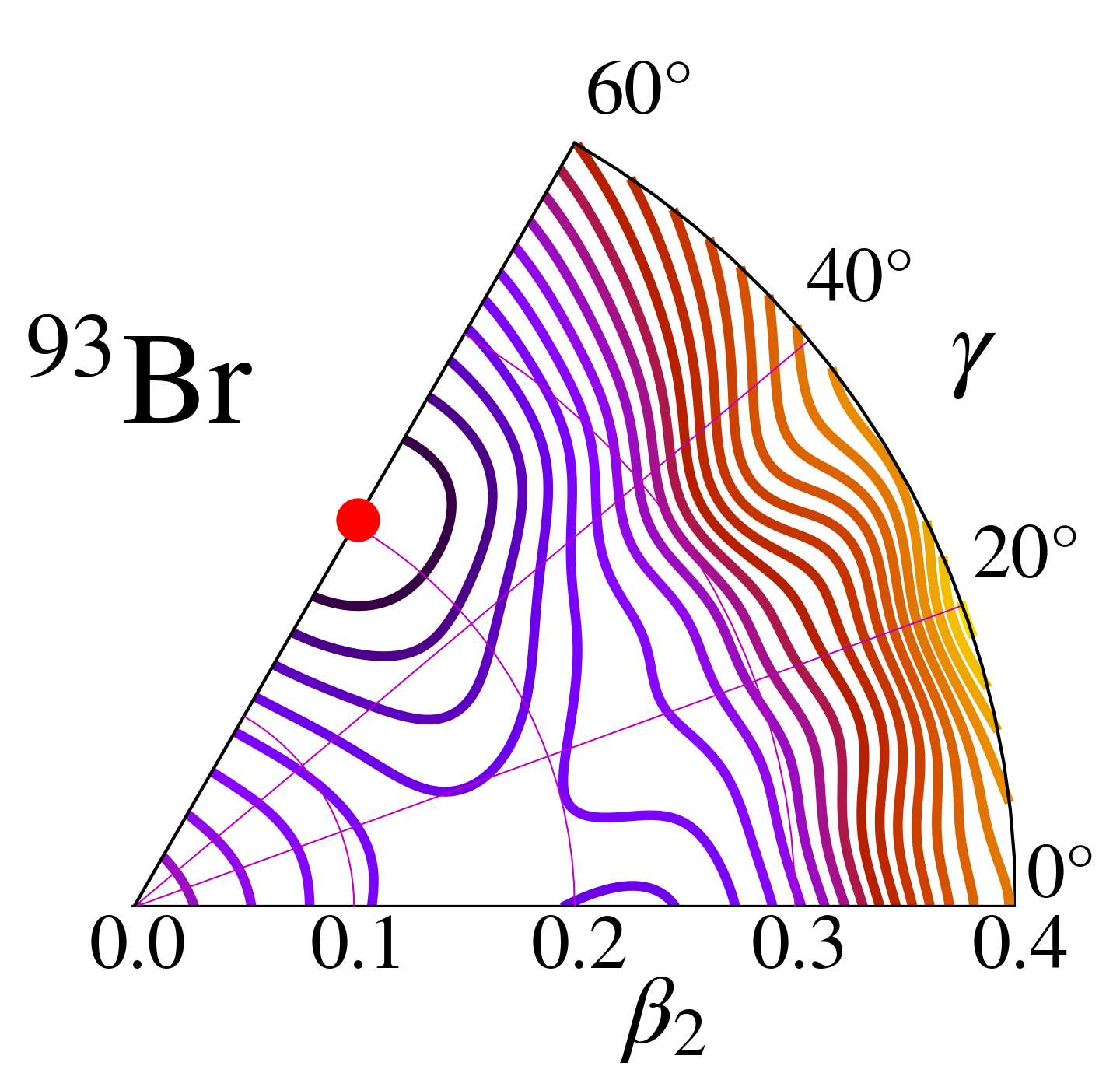}
        \includegraphics[scale=0.32]{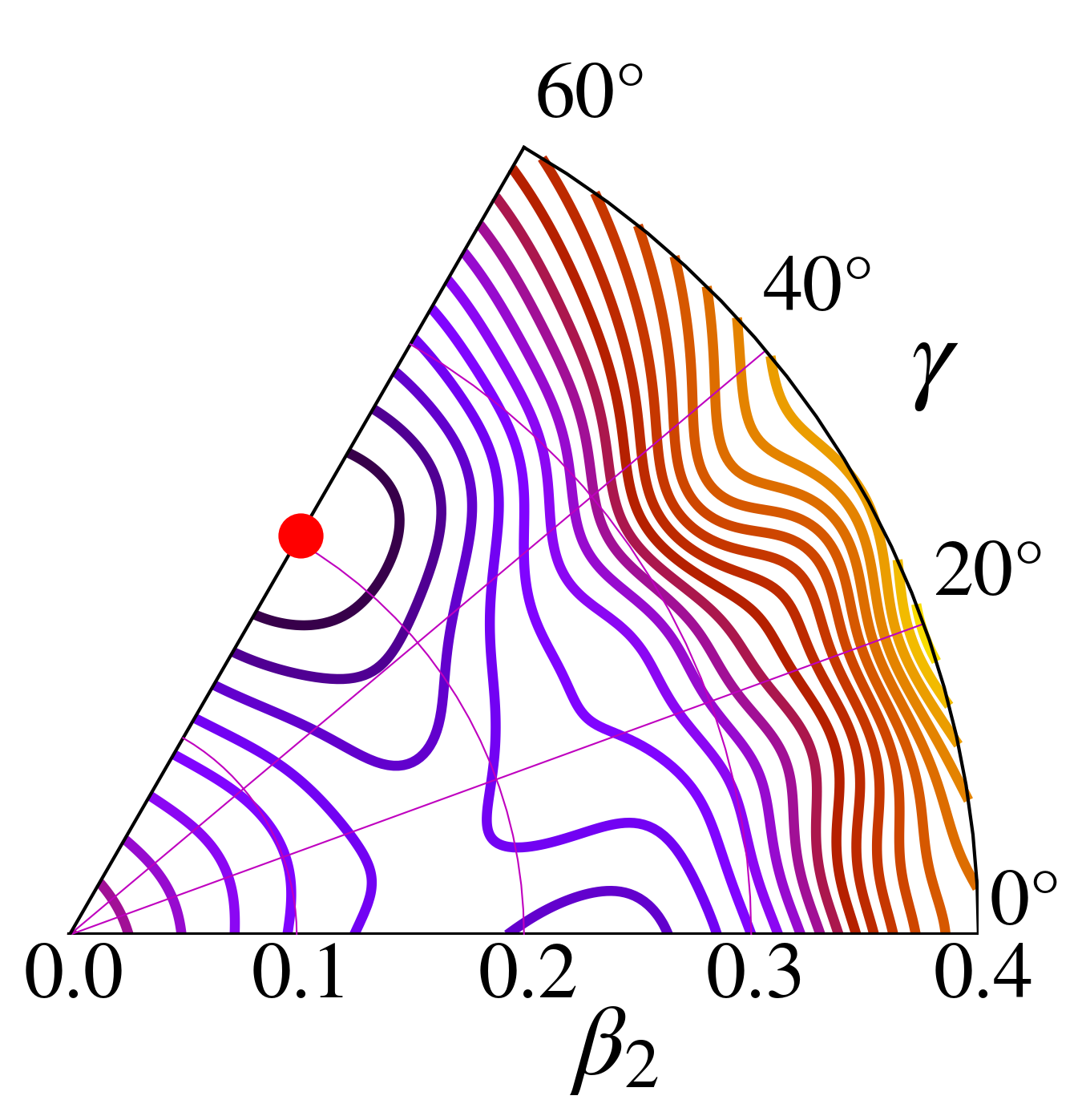}
        \includegraphics[scale=0.32]{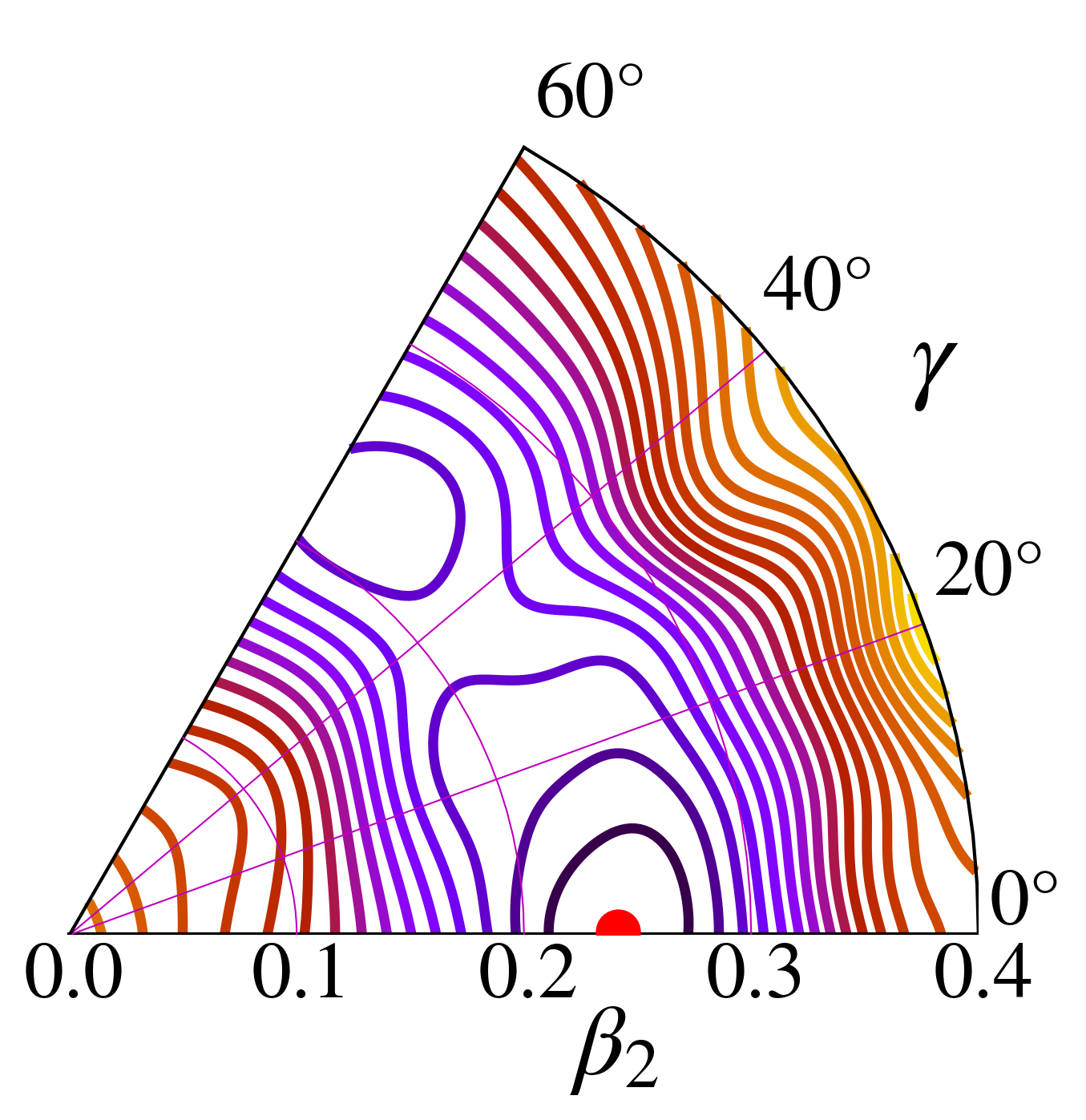}
        \includegraphics[scale=0.32]{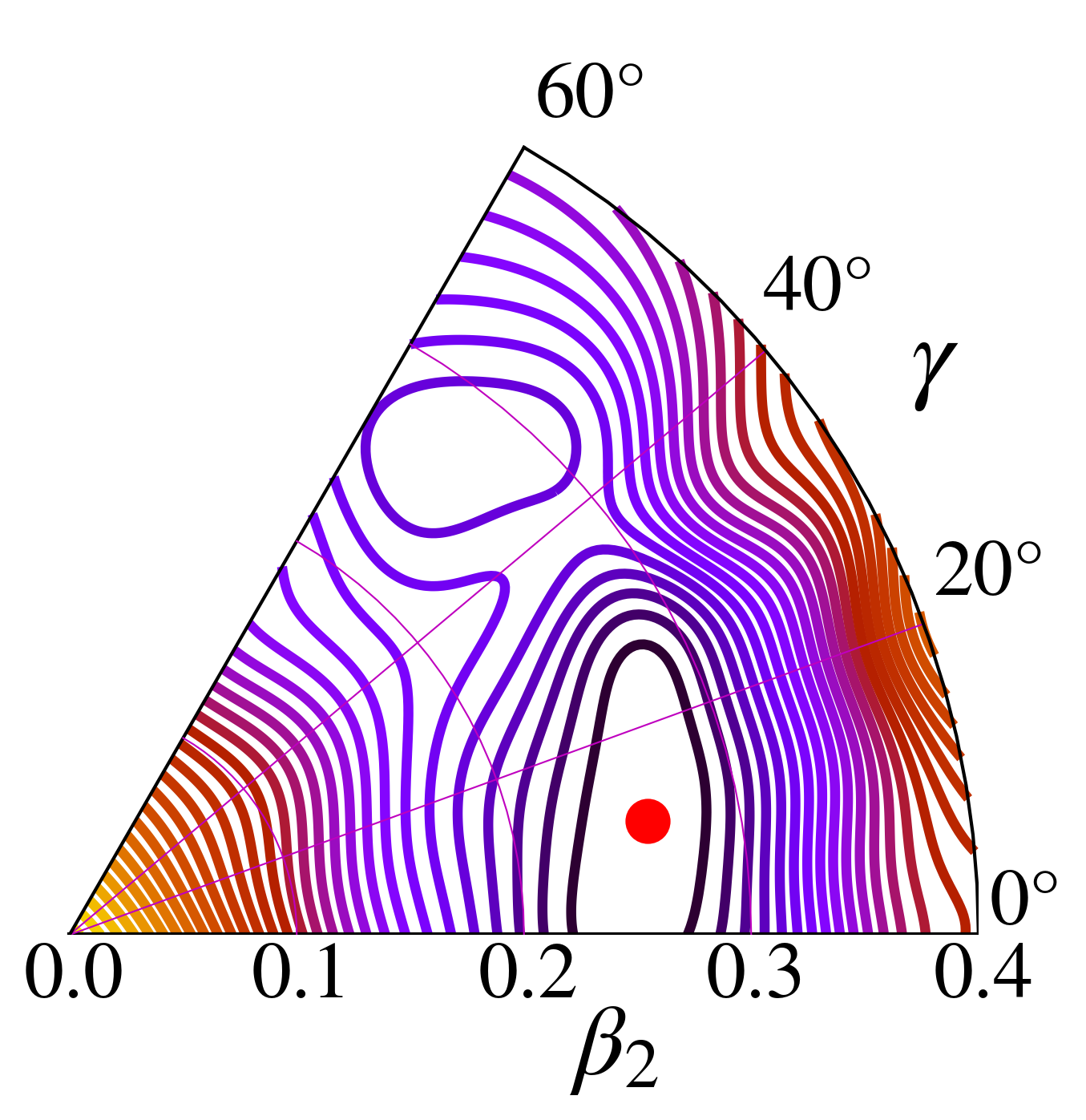}
    \caption{Total Routhian surface plots at four different rotational frequencies, $\omega$ = 0.0, 0.2, 0.4 \& 0.6 (MeV/$\hbar$), in vacuum configuration for $^{85,87,89,91,93}$Br isotopes, respectively. The red-filled circle represents deformation at TRS minimum, and the contour line spacing is $0.2$ MeV.}
    \label{TRS_fig1}
\end{figure*}

TRS provides a comprehensive visualization of the Routhian surface of a nucleus as a function of deformation parameters at a given rotational frequency. These TRS plots depicted in Fig.~\ref{TRS_fig1}, are useful in interpreting the structural evolution and assigning configurations in the present study of $^{85,87,89,91,93}$Br isotopes. 

The $^{85}$Br nucleus with magic neutron number $N = 50$ exhibits a spherical shape at low rotational frequencies \((\omega = 0.0, 0.2)\) due to the strong shell closures that provide extra binding energy and stability. This spherical symmetry directly results from the filled nuclear shells, which minimizes deformation. As the rotational frequency increases, the centrifugal forces acting on the nucleus begin to disrupt this symmetry, gradually favoring deformation to minimize the overall energy. At moderate rotational frequency \((\omega = 0.4)\), the nucleus may remain approximately spherical but begins to show signs of deformation. At a larger rotational frequency \((\omega = 0.6)\), the shape evolves into a slight prolate deformation as the nucleons align along the rotation axis to reduce the rotational energy further. This transition reflects the competition between the shell effects favoring sphericity and the rotational forces inducing deformation. The prolate shape at high frequencies is a hallmark of collective rotational motion, where the alignment of angular momentum plays a dominant role. 

As we transit from the magic neutron number \( N = 50 \) to \( N = 52 \) in \( ^{87} \)Br, the nuclear shape evolves to a prolate deformation. At lower rotational frequencies, the nucleus exhibits gamma softness, characterized by a flat total Routhian energy surface with respect to the triaxial degree of freedom ($\gamma$), allowing the nucleus to explore a range of shapes between prolate and oblate. This gamma softness arises due to the reduced energy cost associated with triaxial deformations. However, as the rotational frequency increases, the gamma softness diminishes, and the nucleus stabilizes in a prolate configuration. Unlike \( ^{85} \)Br, the nucleus \( ^{87} \)Br maintains a relatively stable prolate shape. The persistence of the prolate shape suggests that the additional neutrons beyond the magic number stabilize the deformation, maintaining the same structural configuration across the range of rotational excitations.

Similar to \( ^{87} \)Br, in \( ^{89} \)Br the nucleus initially exhibits gamma softness at low rotational frequencies. However, with increased rotational frequency, the nucleus becomes prolate (\(\beta_2 \sim 0.1\)). At extreme higher frequency \((\omega = 0.6)\), the deformation transitions towards a near oblate shape with other possible minima at triaxial degree \(\gamma \sim 25^\circ \), indicating a significant redistribution of the nuclear structure under the influence of rapid rotation. This evolution reflects the competition between the collective rotational energy and the single-particle effects driving the nuclear shape towards increased stability in different configurations. 

\begin{figure*}[t]
    \centering
    \includegraphics[width=0.98\linewidth]{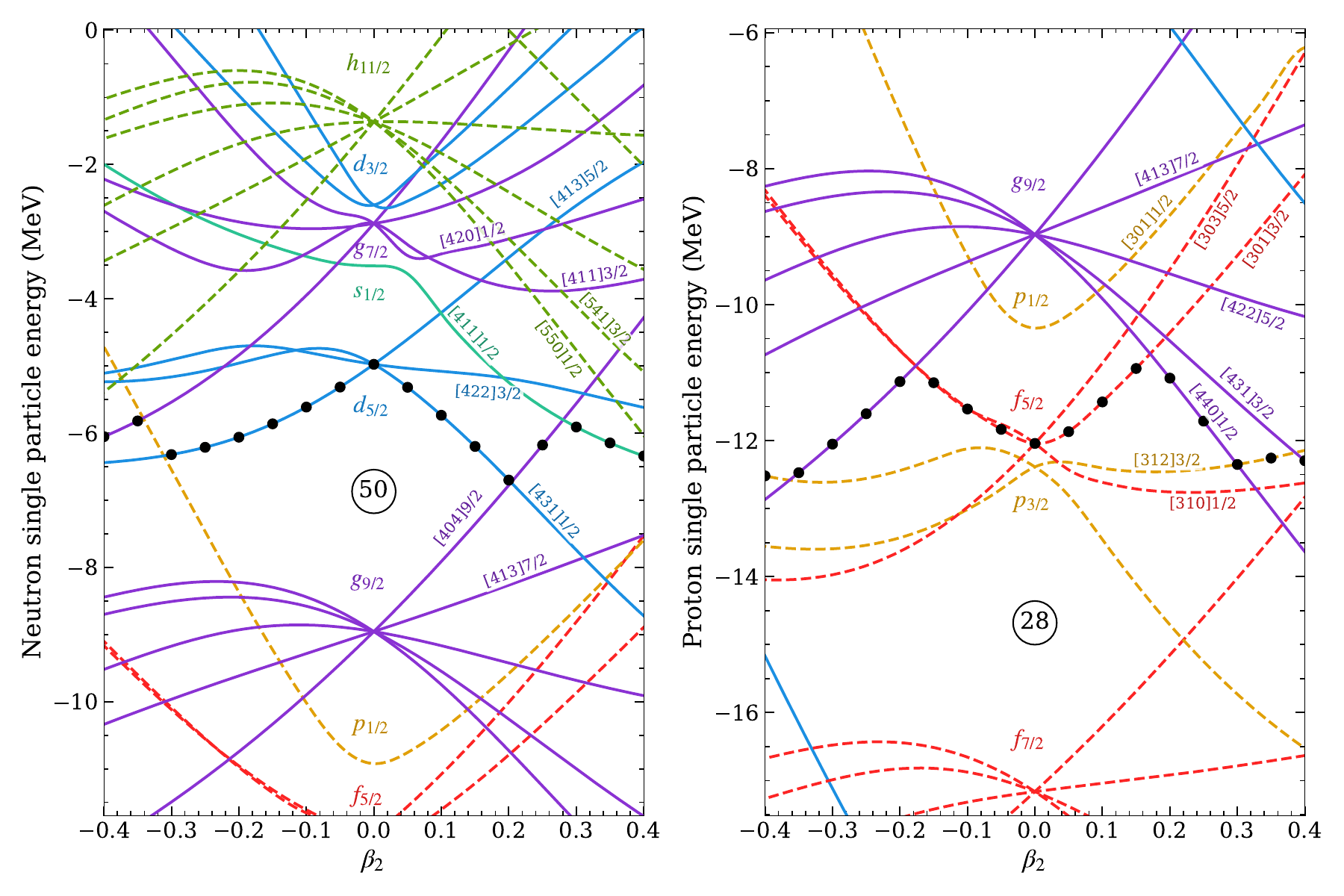}
    \caption{Neutron and proton single-particle levels in $^{87}$Br as a function of the quadrupole deformation $\beta_2$ calculated using the Woods-Saxon potential. Positive (negative) parity lines are represented with solid (dashed) lines. The Fermi level is marked with filled circles. The subshells are indicated for each orbital, together with important asymptotic Nilsson quantum numbers.}
    \label{fig:ws}
\end{figure*}

For \( ^{91} \)Br and \( ^{93} \)Br, the nuclear shape starts as oblate with \(\beta_2 \sim 0.2\) at lower rotational frequencies, contrasting with the prolate shapes observed in \( ^{87,89} \)Br with \(\beta_2 \sim 0.1\). This predicted structural shift from a prolate to oblate transition at $N=56$ is in excellent agreement with recent discrete nonorthogonal (DNO) shell-model calculations~\cite{J.Dudouet}, which identify a prominent pseudo-SU3 driven oblate minimum for these heavier isotopes. As shown in Table~\ref{tab:tab1}, while global macroscopic-microscopic models like the finite range droplet model (FRDM)~\cite{MOLLER1988225} predict persistent prolate shapes, reflecting the broad macroscopic trends of this mass region. However, resolving the localized prolate-to-oblate transition at $N=56$ requires a model that is highly sensitive to specific orbital occupations. By explicitly tracking these single-particle effects, our configuration-constrained CSM successfully reproduces the oblate ground state driven by the evolving neutron Fermi surface.

Furthermore, at the moderate rotational frequency \((\omega = 0.4)\), these nuclei transition back to a prolate configuration, suggesting a stabilization of the shape. However, at \((\omega = 0.6)\), the shape evolves further into a triaxial configuration with another possible minima at an oblate shape. This indicates the interplay between collective rotation and single-particle dynamics, leading to a more complex deformation scenario. This also highlights the sensitive dependence of nuclear shapes on rotational frequency and isotopic structure. The equilibrium quadrupole \( (\beta_2) \) and hexadecapole \( (\beta_4) \) deformations at $\omega = 0.0$ are given in Table~\ref{tab:tab1} along with the finite range droplet model (FRDM)~\cite{MOLLER1988225} values for $^{85,87,89,91,93}$Br.

\begin{table}[t]
\caption{\label{tab:tab1}The quadrupole \( (\beta_2) \) and hexadecapole \( (\beta_4) \) deformation parameters of $^{85,87,89,91,93}$Br in the present work at $\omega = 0.0$ along with FRDM~\cite{MOLLER1988225} values.}
\begin{ruledtabular}
\begin{tabular}{
    l
    S[table-format=-1.3]  
    S[table-format=1.3]   
    S[table-format=-1.3]  
    S[table-format=1.3]   
}
& \multicolumn{2}{c}{CSM} & \multicolumn{2}{c}{FRDM} \\
\cline{2-3} \cline{4-5}
\textrm{Nuclei} & \textrm{$\beta_2$} & \textrm{$\beta_4$} & \textrm{$\beta_2$} & \textrm{$\beta_4$} \\
\colrule
\\[-0.5em]
$^{85}$Br & -0.016 & 0.010 & -0.042 & 0.001 \\
$^{87}$Br & 0.094 & 0.010 & -0.105 & 0.004 \\
$^{89}$Br & 0.115 & 0.011 & 0.195 & 0.002 \\
$^{91}$Br & -0.164 & 0.011 & 0.241 & -0.026 \\
$^{93}$Br & -0.204 & 0.010 & 0.241 & -0.028 \\
\end{tabular}
\end{ruledtabular}
\end{table}

\begin{figure*}[t]
    \includegraphics[width=0.48\linewidth]{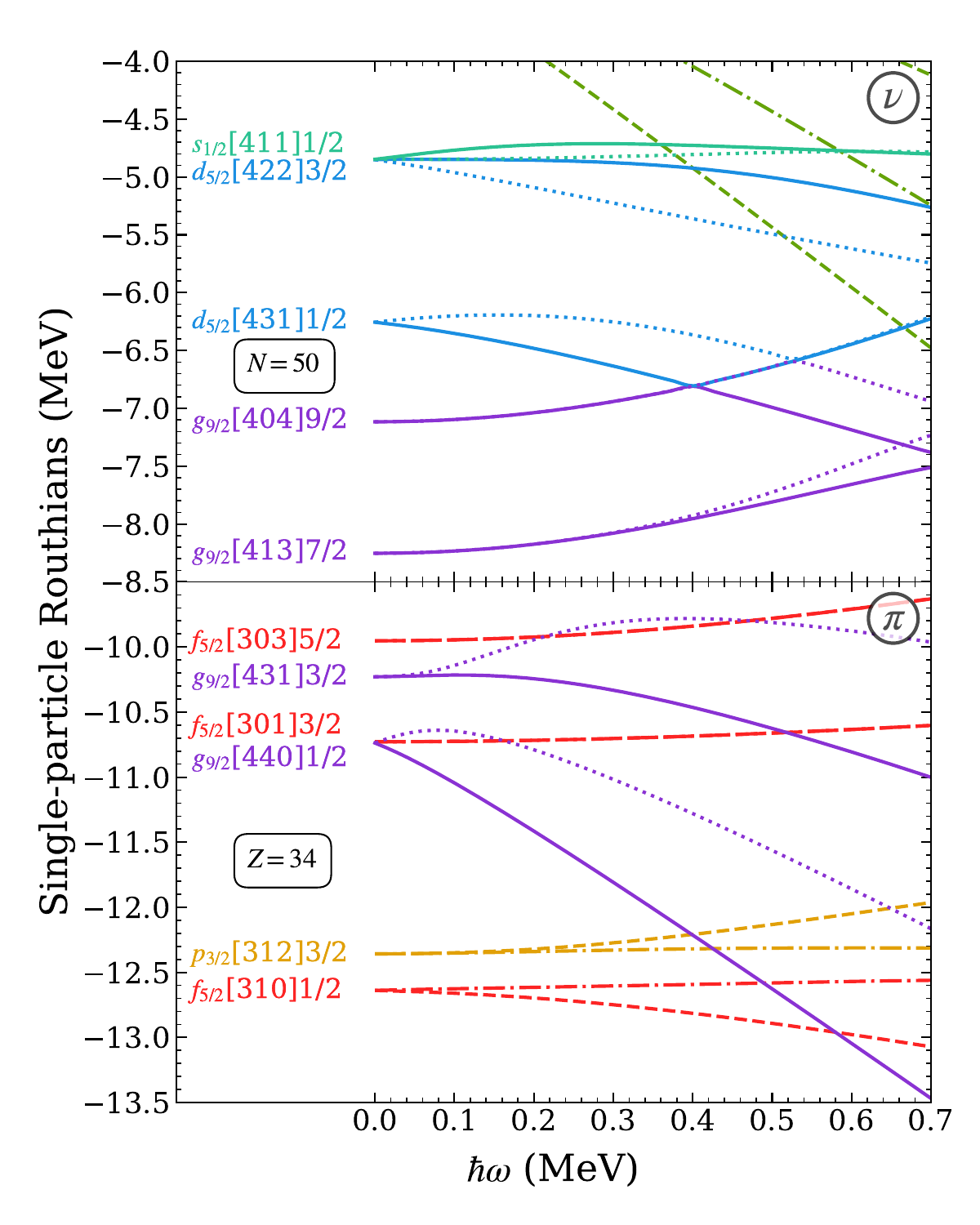}
    \includegraphics[width=0.48\linewidth]{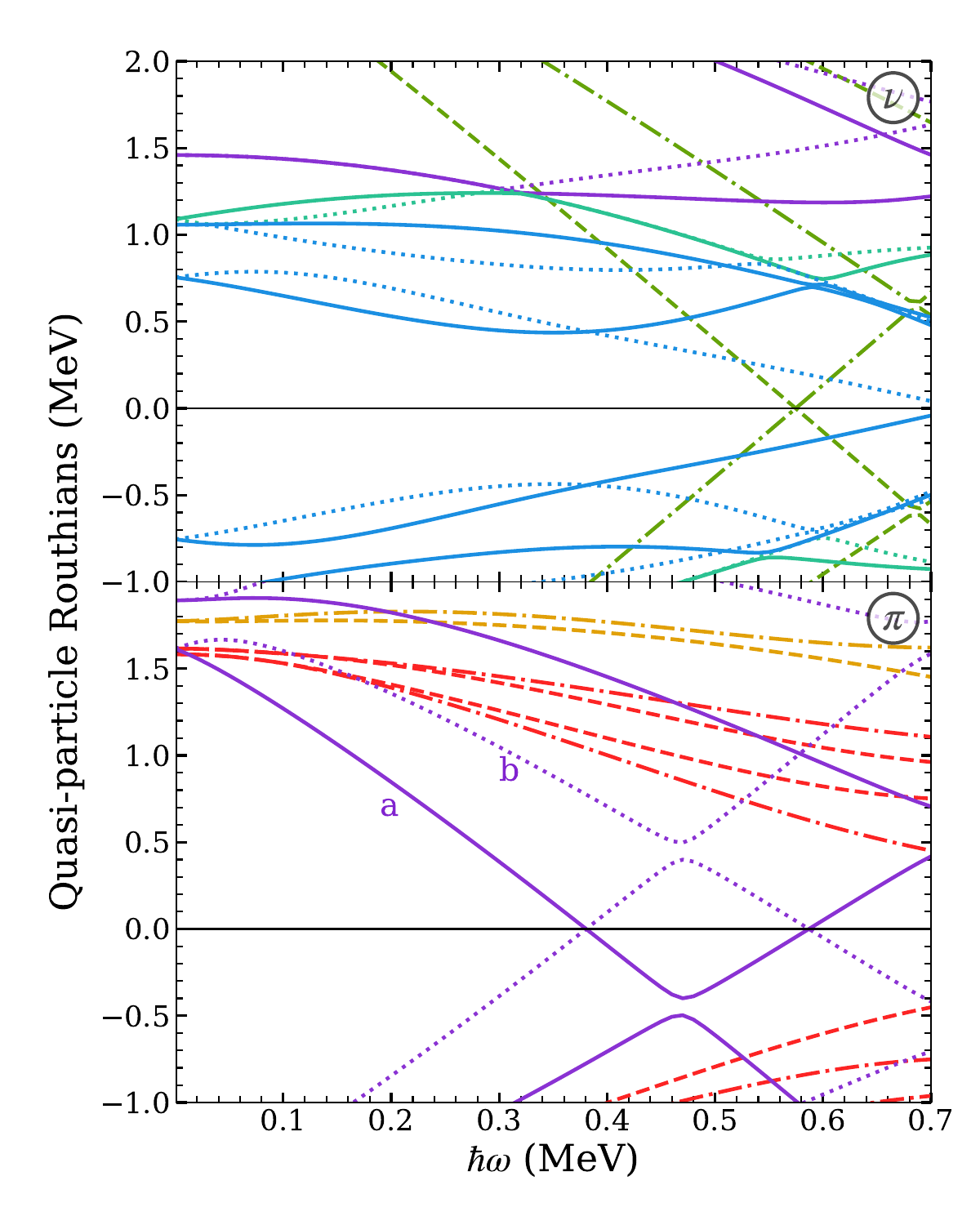}
    \caption{Single-particle Routhians (Left) and quasiparticle Routhians (Right) for neutrons $(\nu)$ and protons $(\pi)$ close to the Fermi level of $^{87}$Br as a function of rotational frequency for $\beta_2 =0.1668$. Parity and signature $(\pi,\alpha)$ are represented by solid $(+,+1/2)$, dotted $(+,-1/2)$, dot-dash $(-,+1/2)$ and dashed $(-,-1/2)$ lines.}
    \label{SPR_QPR}
\end{figure*}

The single-particle level scheme shown in Fig.~\ref{fig:ws} illustrates the evolution of neutron and proton orbitals in $^{87}$Br as a function of quadrupole deformation $\beta_2$. This representation serves as a reference for the entire isotopic chain, since the overall structure of the Nilsson diagram remains similar for neighboring Br nuclei, with only shifts in the energy levels and the Fermi level depending on neutron number. The orbitals lying close to the Fermi level are of particular importance, as they dictate the available quasiparticle configurations relevant for the low-lying rotational structures discussed later in this work. The asymptotic Nilsson quantum numbers indicated in the figure help identify these key orbitals and provide guidance for configuration assignments, especially in regions where band crossings and alignments play a role in shaping the observed rotational behavior.

In Fig.~\ref{SPR_QPR}, the single-particle Routhians and quasi-particle Routhians are shown for neutrons and protons near the Fermi surface of \(^{87}\)Br. The structure of single-particle Routhians of \(^{87,89,91,93}\)Br are similar in trend, which are shown for \(^{87}\)Br at \(\beta_2 = 0.1668\) deformation, corresponding to one-quasiproton configuration `a' explained in Appendix~\ref{Appendix_A}. The energy levels in deformed nuclei are represented based on their symmetry properties, parity (\(\pi\)) and signature (\(\alpha\)) using different line styles and labelled with letters as detailed in Table~\ref{tab:tab2}. 

The alignment and interaction of single-particle states change as nucleon energy evolves with increasing rotational frequency. To study these changes, single-particle Routhians are examined and classified using the Nilsson states labelled with the quantum numbers \( \Omega [N n_z \Lambda] \)~\cite{nilsson1955}. As \(\omega\) increases, the states may mix, but the original quantum numbers indicate their starting point. Most trajectories remain relatively unchanged as \(\omega\) increases, but some show significant differences within the displayed frequency range. A quasiparticle Routhian diagram provides a graphical representation of quasiparticle energy levels as a function of the rotational frequency, with each trajectory corresponding to a specific quasiparticle state. As an example, \( N=52 \) the valence particle will occupy the first level above the zero of the quasi-particle Routhian.  This level changes drastically with rotation, ranging from the $g_{9/2}$ to the $h_{11/2}$. Further, we can observe points where energy levels intersect or come close together, indicating band crossings, which are significant for angular momentum alignment. 

\begin{table}[t]
\begin{ruledtabular}
    \caption{\label{tab:tab2}The representation of quasiparticle levels with different combinations of parity \( (\pi) \) and signature \( (\alpha) \).}
    \centering
    \begin{tabular}{cccc}
          Sr. No. & Parity ($\pi$) & Signature ($\alpha$)  & Label  \\ \hline
         1. & $-$ & $-1/2$ & e, g, m, o \\ 
         2. & $-$ & $+1/2$ & f, h, n, p \\ 
         3. & $+$ & $-1/2$ & b, d, j, l \\ 
         4. & $+$ & $+1/2$ & a, c, i, k \\ 
    \end{tabular}
\end{ruledtabular}
\end{table}

\begin{figure*}[t]
    \includegraphics[scale=0.65]{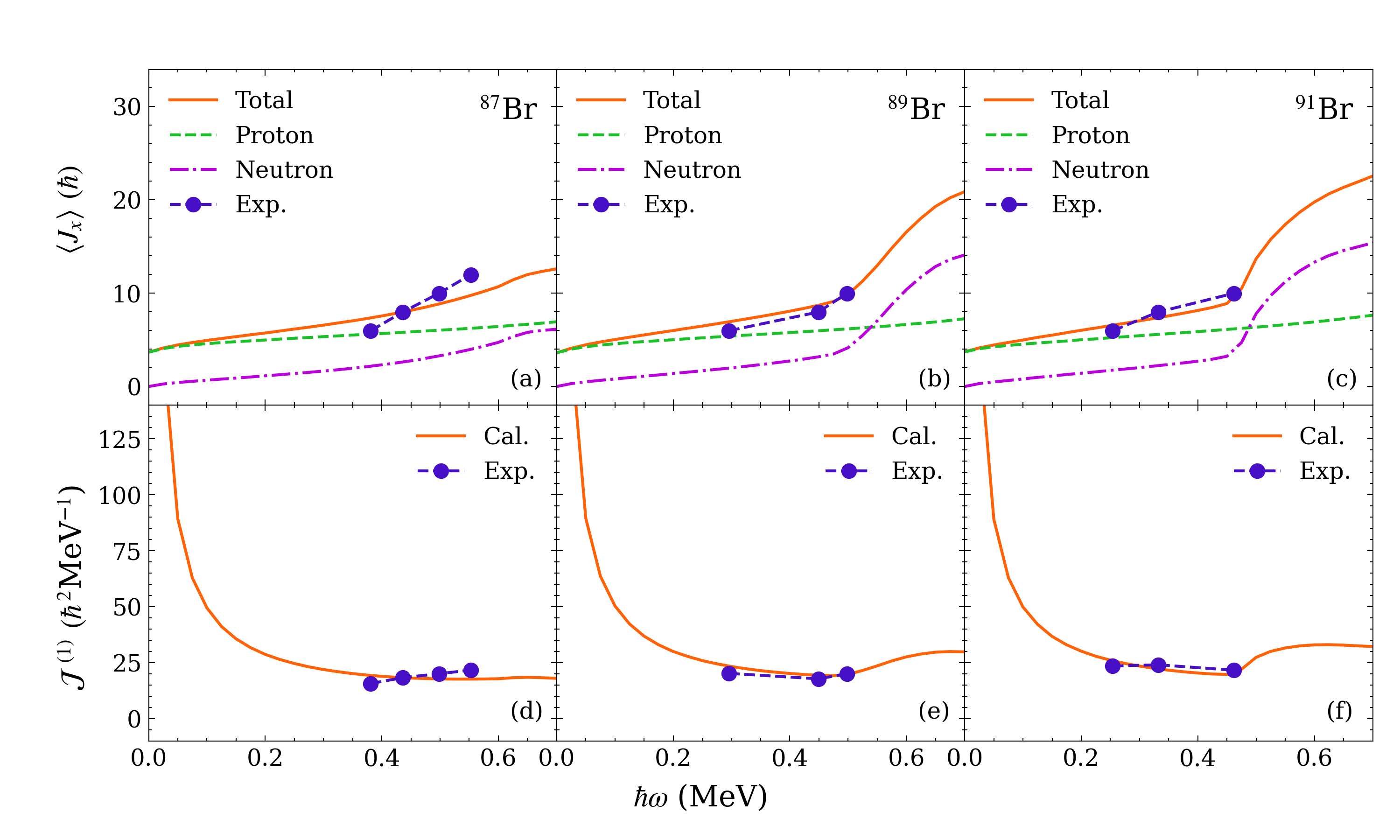}
    \caption{Aligned angular momentum $\langle J_x \rangle$ and kinematic moment of inertia $\mathcal{J}^{(1)}$ for Band~1 in $^{87,89,91}\mathrm{Br}$, calculated using the CSM (solid line). The dashed and dash--dotted lines denote the proton and neutron contributions, respectively. Experimental data are shown by blue-filled circles.}
\label{jx_moi}
\end{figure*}
\begin{figure}[t]
    \centering
    \includegraphics[scale=0.37]{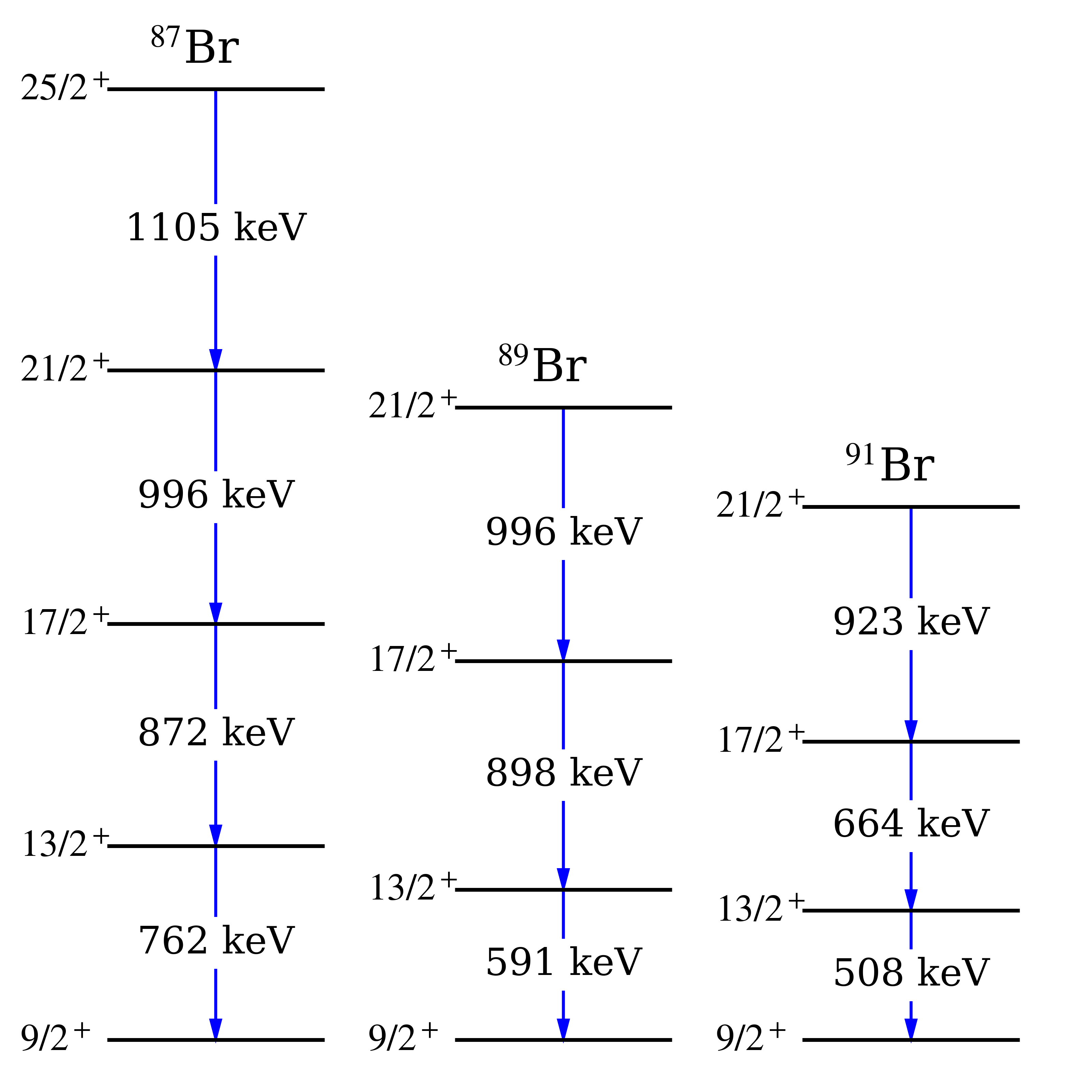}
    \caption{A comparison of the experimentally deduced positive parity band 1 in $^{87,89,91}$Br isotopes~\cite{J.Dudouet}.}
    \label{nls}
\end{figure}

\subsection{Aligned angular momentum and KMoI}
\label{subsec:Aligned}
In \(^{87}\)Br, the positive-parity rotational band with bandhead spin \(I^\pi=9/2^+\) is assigned to the one-quasiproton configuration `a', given in Table~\ref{tab:tab2}, based on the \(\pi g_{9/2}[440]1/2^+\) Nilsson orbital. This assignment is supported by the location of the orbital near the proton Fermi surface in the Routhian diagrams (Fig.~\ref{SPR_QPR}) and by the experimental level energies and transition pattern displayed in Fig.~\ref{nls}~\cite{J.Dudouet}. The sequence provides a well-defined example of a decoupled high-\(j\) proton band built on a non-yrast configuration~\cite{Br87paper}.

The experimental rotational frequency \(\hbar\omega(I)\) used for the alignment analysis is extracted from the measured \(\gamma\)-transition energies using Eq.~\eqref{eq:exp_w}. The theoretical aligned angular momentum \(\langle J_x\rangle\) and kinematic moment of inertia \(\mathcal{J}^{(1)}\) are obtained from Eqs.~\eqref{eq:jx} and \eqref{eq:kmoi}, respectively. The results, shown in Fig.~\ref{jx_moi}\textcolor{blue}{(a)}, demonstrate that the proton contribution to \(\langle J_x\rangle\) remains nearly constant across the frequency range due to the blocking of the \(\pi g_{9/2}\) orbital, whereas the observed increase in total alignment is driven almost entirely by neutron quasiparticles close to the Fermi surface, as supported by the quasiparticle Routhians in Fig.~\ref{SPR_QPR}. This highlights the crucial role of the even-\(N\) core in generating angular momentum, despite the band being structurally defined by a single odd proton, a characteristic feature of odd-\(Z\), even-\(N\) nuclei in this mass region~\cite{bengtsson1979,Frauendorf2001}.

The measured band shows the typical characteristics of a decoupled sequence built on a deformation-driving orbital: relatively large \(\mathcal{J}^{(1)}\) shown in Fig.~\ref{jx_moi}\textcolor{blue}{(d)}, weak signature splitting, and a prolate equilibrium shape consistent with the TRS minimum at \(\beta_2\approx0.1668\) as given in Table~\ref{tab:tab3} of Appendix~\ref{Appendix_A} . The band lies \(\sim 1\) MeV above the yrast line and only one signature partner is firmly observed, indicating its non-yrast character and reduced competition from yrast mixing. These features are consistent with previous systematic studies of \(g_{9/2}\) configurations in neighboring nuclei~\cite{Br87paper}. While the cranked shell model reproduces the alignment at the bandhead, it systematically underestimates \(\langle J_x\rangle\) at higher frequencies, with the gap increasing toward \(\hbar\omega \sim 0.6\) MeV. Possible contributors to this deviation include progressive reduction of neutron pairing with increasing frequency and dynamic core polarization effects from the deformation-driving \(\pi g_{9/2}\) orbital. The overall picture is that the \(9/2^+\) band in \(^{87}\)Br represents a blocked-\(\pi g_{9/2}\) configuration in which the odd proton fixes the underlying structure while the alignment is largely carried by neutrons. The discrepancy between theory and experiment highlights the importance of neutron-driven collectivity and evolving correlations, pointing to the need for improved treatments of pairing reduction and core polarisation effects in non-yrast bands. 

The alignment behavior observed in \(^{89}\)Br and \(^{91}\)Br exhibits a remarkably similar trend, with both nuclei showing a gradual increase in the aligned angular momentum as the rotational frequency rises. The results, illustrated in Fig.~\ref{jx_moi}\textcolor{blue}{(b)} for \(^{89}\)Br and in Fig.~\ref{jx_moi}\textcolor{blue}{(c)} for \(^{91}\)Br indicate that the proton contribution to \(\langle J_x \rangle\) remains essentially unchanged throughout the rotational-frequency range, consistent with the blocking of the \(\pi g_{9/2}\) orbital. In contrast, the gradual increase in total alignment arises predominantly from neutron quasiparticles. This trend closely parallels the alignment mechanism observed in \(^{87}\)Br, reflecting the systematic role of neutron-driven collectivity within this isotopic chain.
The configuration chosen for the calculations is a one-quasiproton configuration, labelled `a' in Table~\ref{tab:tab2}. Within the limited frequency ranges where experimental data is currently available ($\hbar\omega \approx 0.27-0.55$ MeV for $^{89}$Br and $0.25-0.5$ MeV for $^{91}$Br), our calculated aligned angular momentum $\langle J_x \rangle$ and KMoI \( \mathcal{J}^{(1)} \) are in excellent agreement with the measured values. Furthermore, our CSM results in this low-to-moderate spin regime are highly comparable to recent state-of-the-art large scale shell model (LSSM) calculations~\cite{J.Dudouet}.

However, at higher rotational frequencies ($\hbar\omega \approx 0.55$ MeV for $^{89}$Br and $0.48$ MeV for $^{91}$Br), the theoretical calculations predict a sharp upbend in the aligned angular momentum $\langle J_x \rangle$ beyond the $21/2^+$ state. Consequently, the kinematic moments of inertia $\mathcal{J}^{(1)}$, shown in Figs.~\ref{jx_moi}\textcolor{blue}{(e)} and \ref{jx_moi}\textcolor{blue}{(f)}, transition from a relatively flat plateau characteristic of rotational bands prior to a major alignment to a corresponding abrupt increase. Because our calculation rigorously constrains the odd proton to the $\pi g_{9/2}$ configuration, this high-spin feature is unambiguously driven by the underlying even-$N$ core. It indicates a multi-quasiparticle band crossing caused by the rotational alignment of a pair of core neutrons. Furthermore, because the pairing field is treated self-consistently using the BCS approach at each frequency step, this alignment is accompanied by a dynamic reduction in the neutron pairing gap due to the Coriolis anti-pairing effect.

Crucially, this predicted core crossing occurs just above the currently observed experimental limits. The lack of experimental data at these frequencies naturally aligns with the physics of pair-breaking. Which is the significant energetic cost required to break the neutron pair, coupled with the resulting fragmentation of $\gamma$-decay intensity. This often prevents these aligned states from being distinctly populated in recent fission experiments. As we move from $^{87}$Br to $^{91}$Br, the observed structural variations and the exact crossing frequencies arise primarily from the changing neutron Fermi surface. These systematic similarities provide a useful basis for interpreting the rotational properties of odd-$Z$ nuclei in this mass region and motivate future experiments aimed at extending measurements toward higher-spin states.

\section{\label{sec:conclu}Conclusions}
We have performed configuration-constrained cranked shell model calculations for $^{87,89,91}$Br isotopes, analyzing total Routhian surfaces, alignments, and kinematic moments of inertia. The TRS results reveal a clear evolution of nuclear shapes along the isotopic chain: $^{85}$Br remains spherical at low rotational frequencies due to the $N=50$ shell closure, while $^{87,89}$Br exhibit prolate shapes with gamma softness at low frequencies. In $^{91,93}$Br, oblate shapes appear at low frequency, transitioning to triaxial configurations at higher rotational frequencies. These ground-state shape transitions are in excellent agreement with recent discrete nonorthogonal (DNO) shell-model calculations~\cite{J.Dudouet}.

Single-particle and quasiparticle Routhians illustrate the role of orbitals near the Fermi surface in driving alignment. Furthermore, the assignment of specific band configurations is rigorously justified in this work; by correlating our theoretical predictions with experimental spectra, we established that the low-lying positive-parity sequences are firmly built upon the decoupled one-quasiproton $\pi g_{9/2}[440]1/2^+$ configuration. In these bands, the structural foundation is defined by the blocked proton, while the rotational collectivity is dominated by the alignment of neutron quasiparticles, a decoupled feature consistently observed across $^{87,89,91}$Br.

Calculated aligned angular momentum $\langle J_x \rangle$ and kinematic moments of inertia $\mathcal{J}^{(1)}$ remarkably reproduce experimental trends up to the currently observed experimental limits (e.g., the $21/2^+$ states in $^{89,91}$Br). Our theoretical calculations extend beyond these existing data points to predict an up-bend at higher rotational frequencies ($\hbar\omega > 0.45$ MeV). This feature indicates a major structural shift driven by a multi-quasiparticle band crossing, likely the alignment of a pair of $g_{9/2}$ neutrons. The emergence of this theoretical crossing provides a compelling physical explanation for the current experimental boundaries, as the energetic cost of pair breaking and the subsequent fragmentation of the $\gamma$-decay flux likely inhibit the population of these higher-spin states in fission experiments.

In summary, our study highlights the sensitivity of nuclear shapes and rotational behavior to neutron number and rotational frequency. The results demonstrate the predictive power of the cranked shell model in capturing dynamic high-spin features. This provides a coherent picture of structural evolution in odd-$Z$ bromine isotopes and serves as a strong motivation for future high-statistics experiments aimed at probing these high-spin frontiers.

\begin{acknowledgments}
We acknowledge the National Supercomputing Mission (NSM) for providing the computing resources of ‘PARAM Ganga’ at the
Indian Institute of Technology Roorkee. This
support is implemented by C-DAC and funded
by the Ministry of Electronics and Information
Technology (MeitY) and the Department of
Science and Technology (DST), Government
of India (GoI). This work is also supported by the Science and Engineering Research Board (SERB) under Grant Code: CRG/2022/009359.
\end{acknowledgments}

\appendix

\section{Configuration-constrained TRS}
\label{Appendix_A}
The configuration-constrained total Routhian surface (TRS) calculations are performed to investigate the equilibrium deformation properties of $^{87,89,91}$Br isotopes for the selected quasiparticle configuration labeled as configuration `a'. In this approach, specific quasiparticle orbitals are blocked to preserve the underlying microscopic configuration while minimizing the total Routhian energy in the $(\beta_2,\gamma)$ deformation plane at each rotational frequency.

\begin{figure}[tb]
\centering

\includegraphics[width=0.32\linewidth]{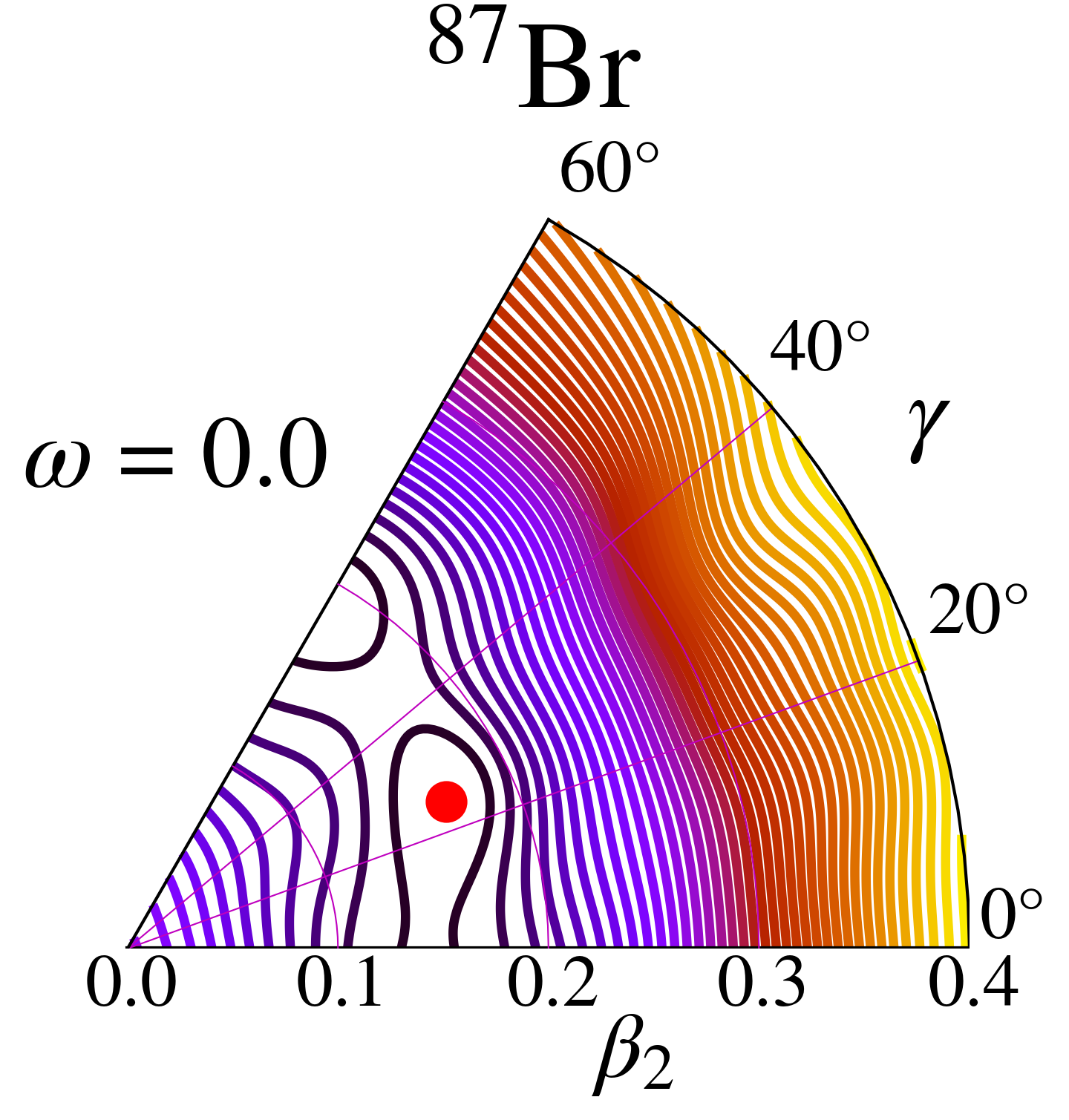}
\includegraphics[width=0.32\linewidth]{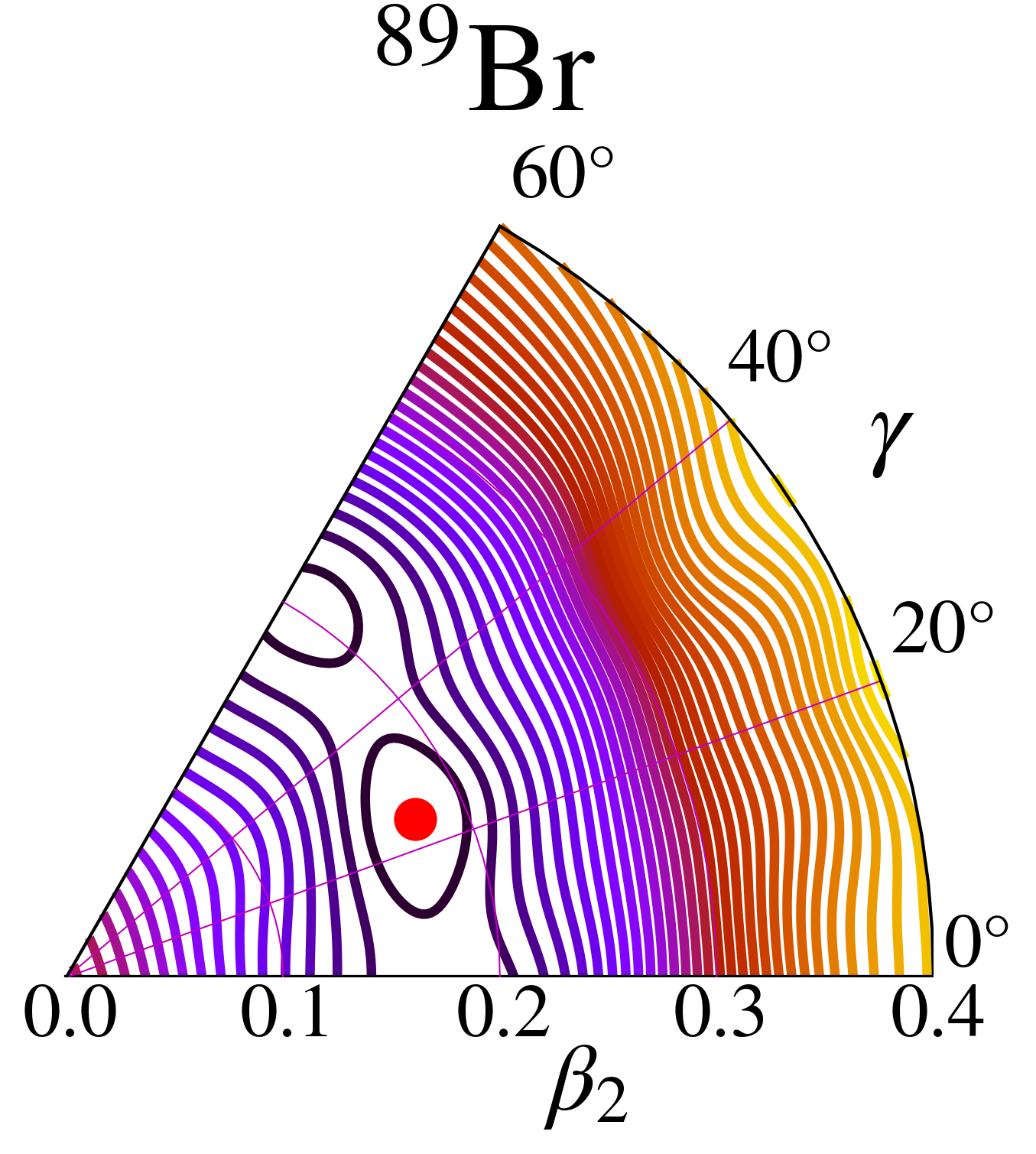}
\includegraphics[width=0.32\linewidth]{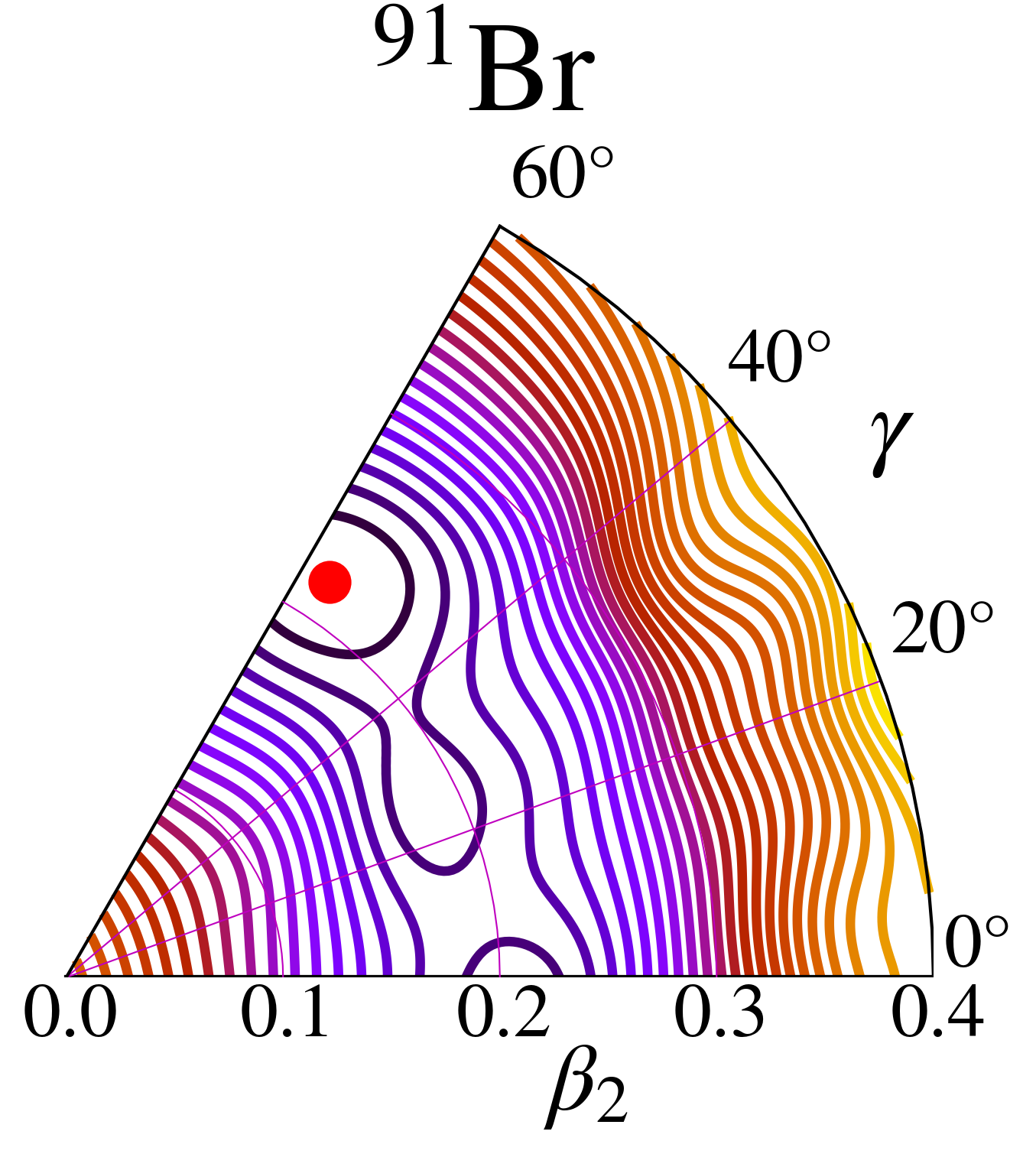}

\includegraphics[width=0.32\linewidth]{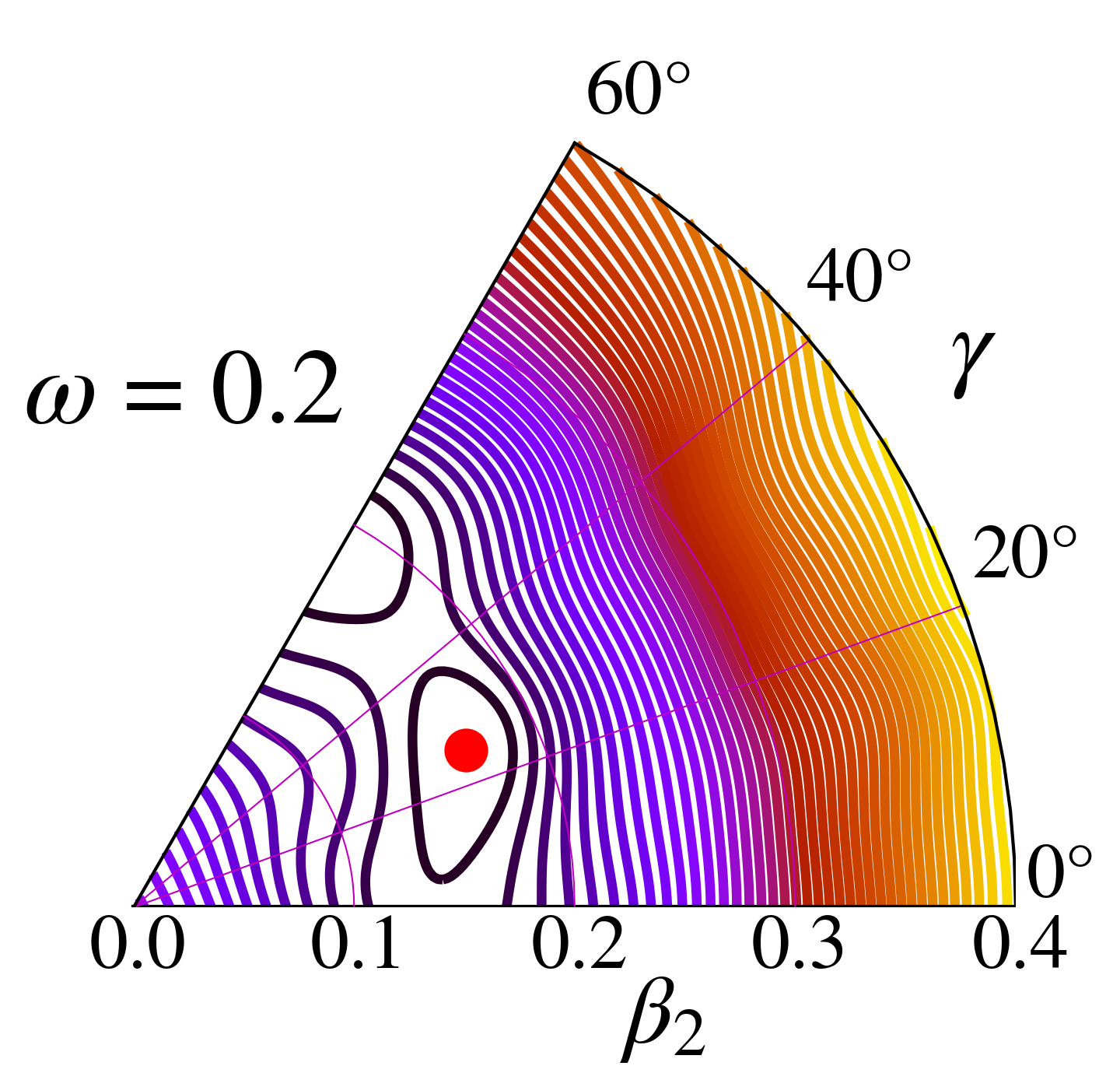}
\includegraphics[width=0.32\linewidth]{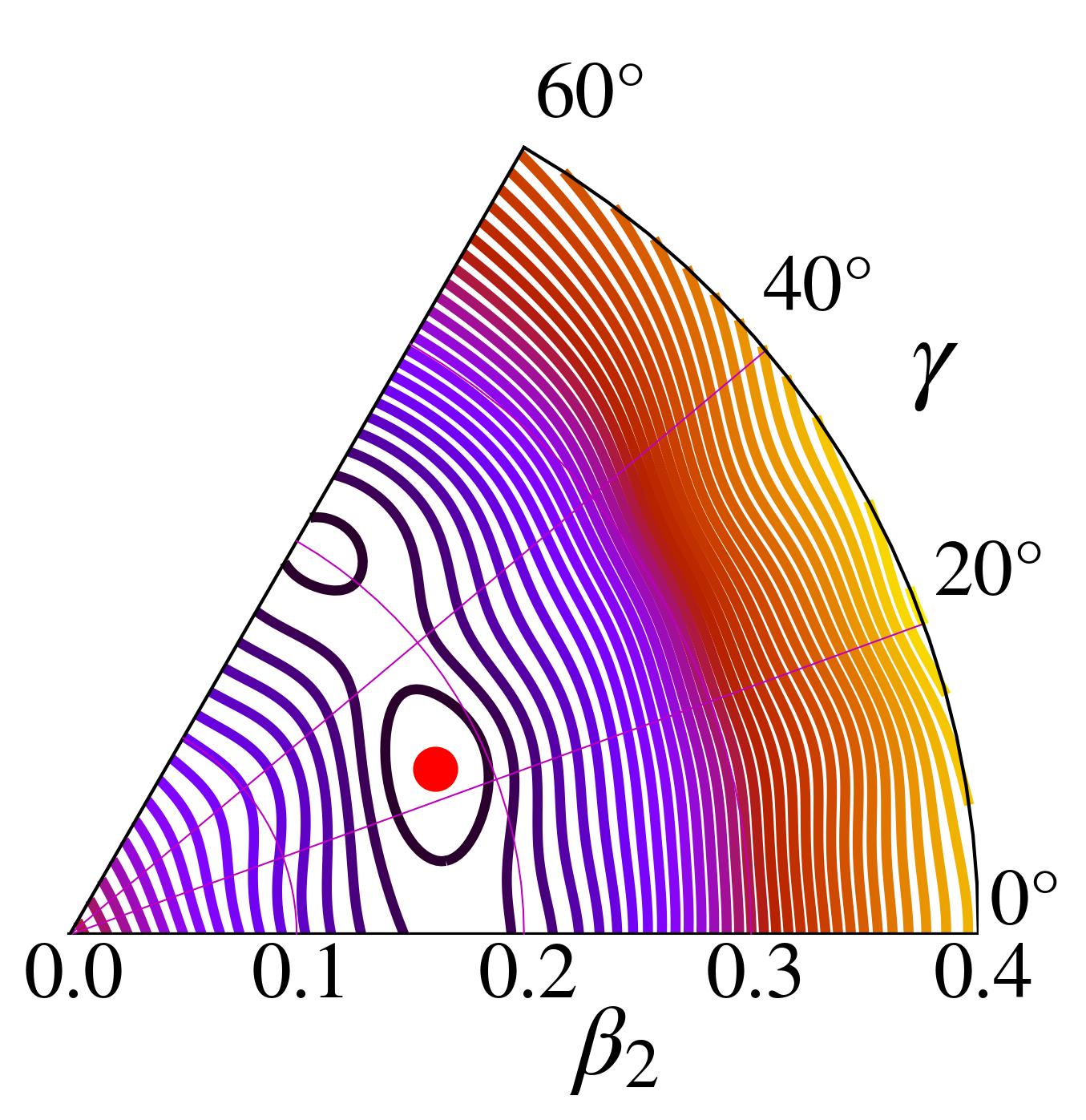}
\includegraphics[width=0.32\linewidth]{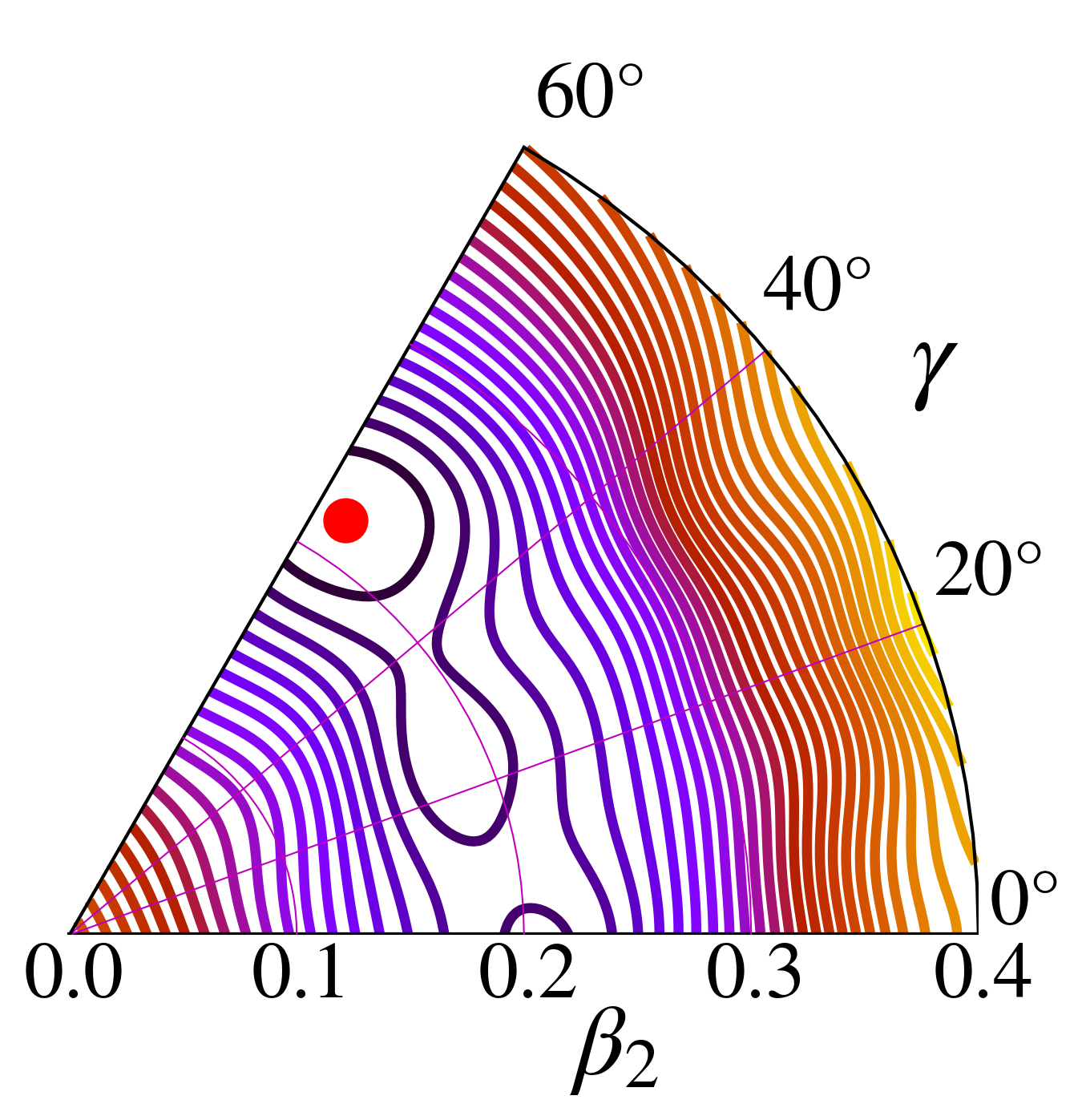}

\includegraphics[width=0.32\linewidth]{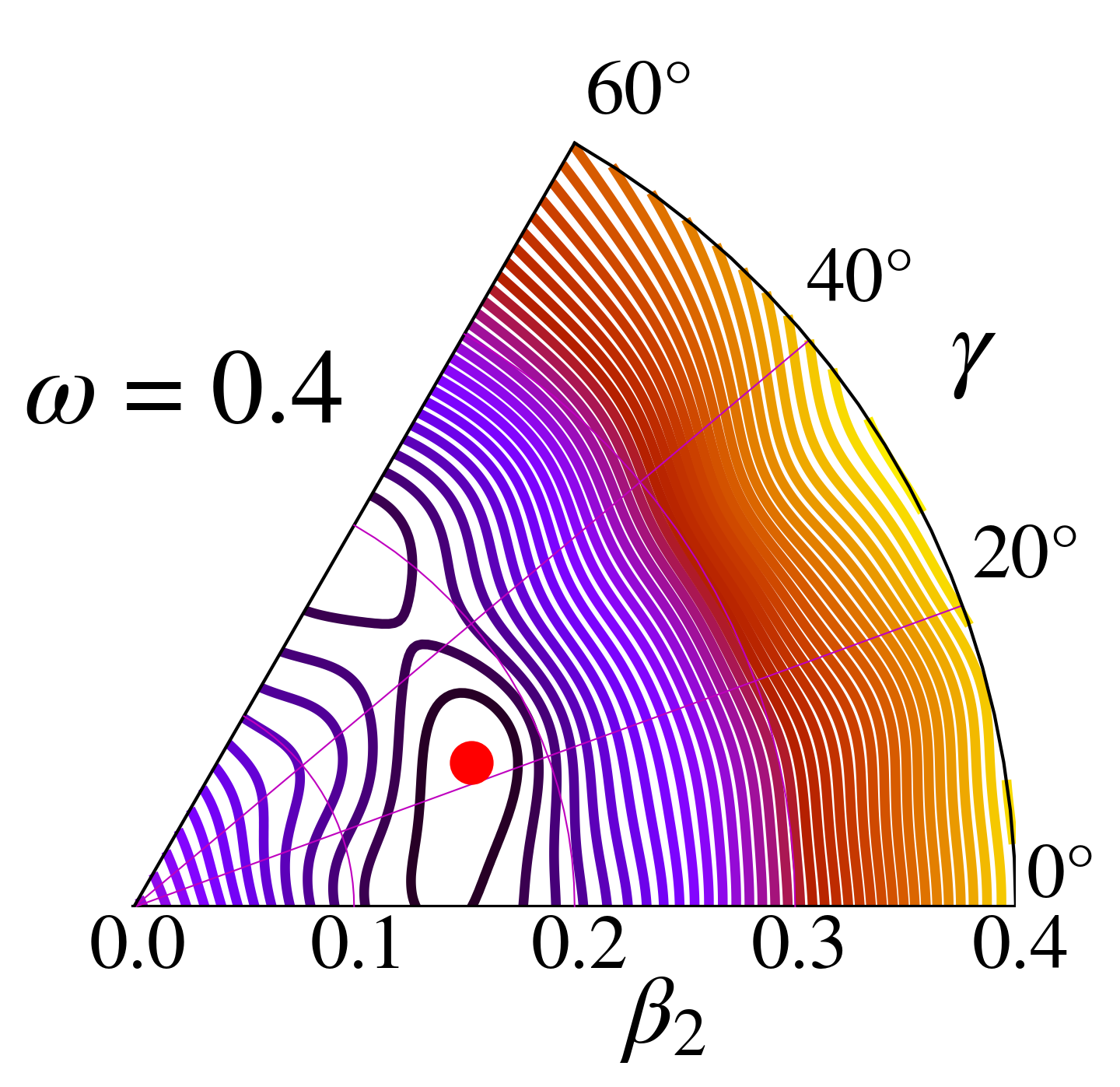}
\includegraphics[width=0.32\linewidth]{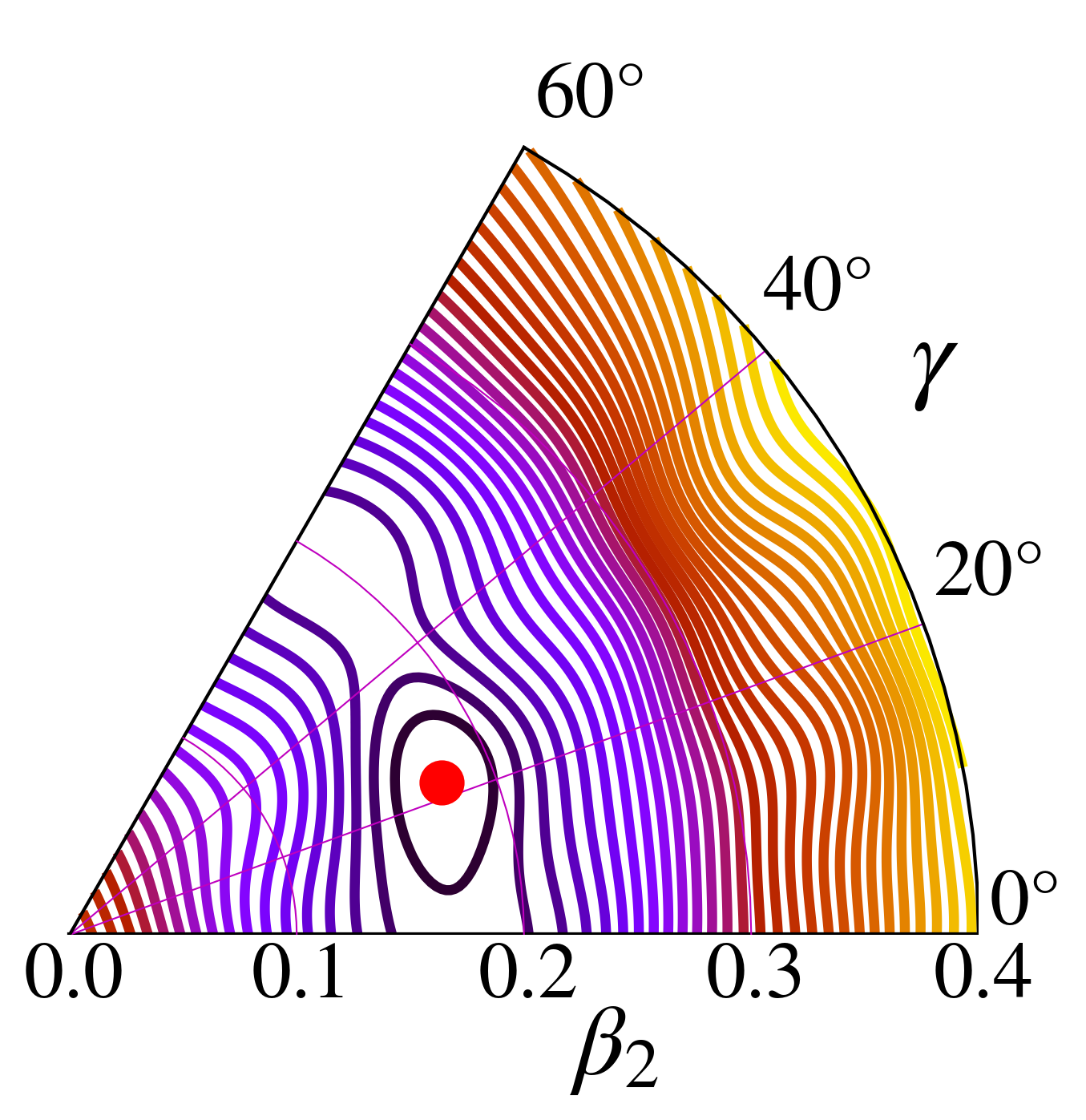}
\includegraphics[width=0.32\linewidth]{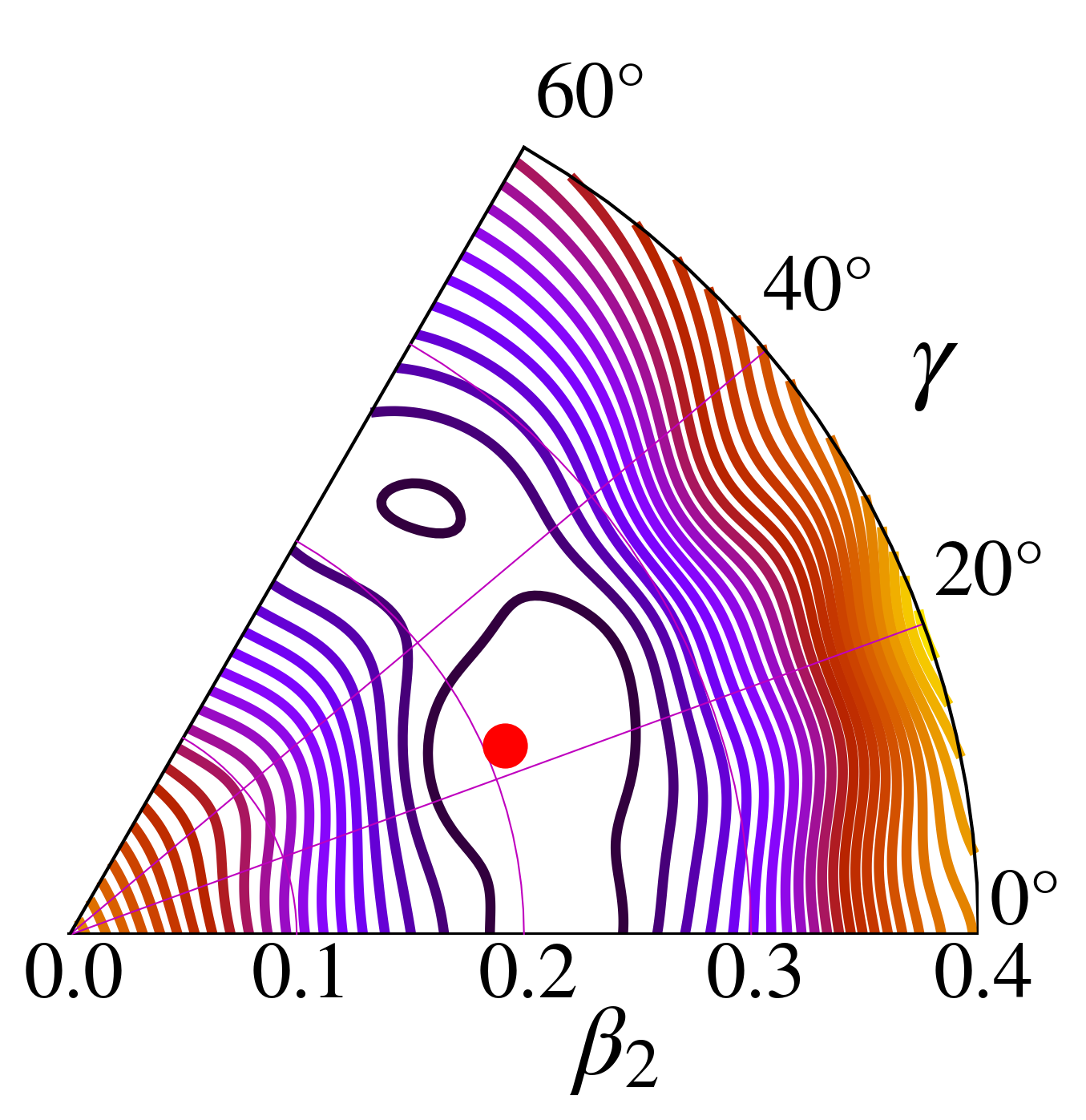}

\includegraphics[width=0.32\linewidth]{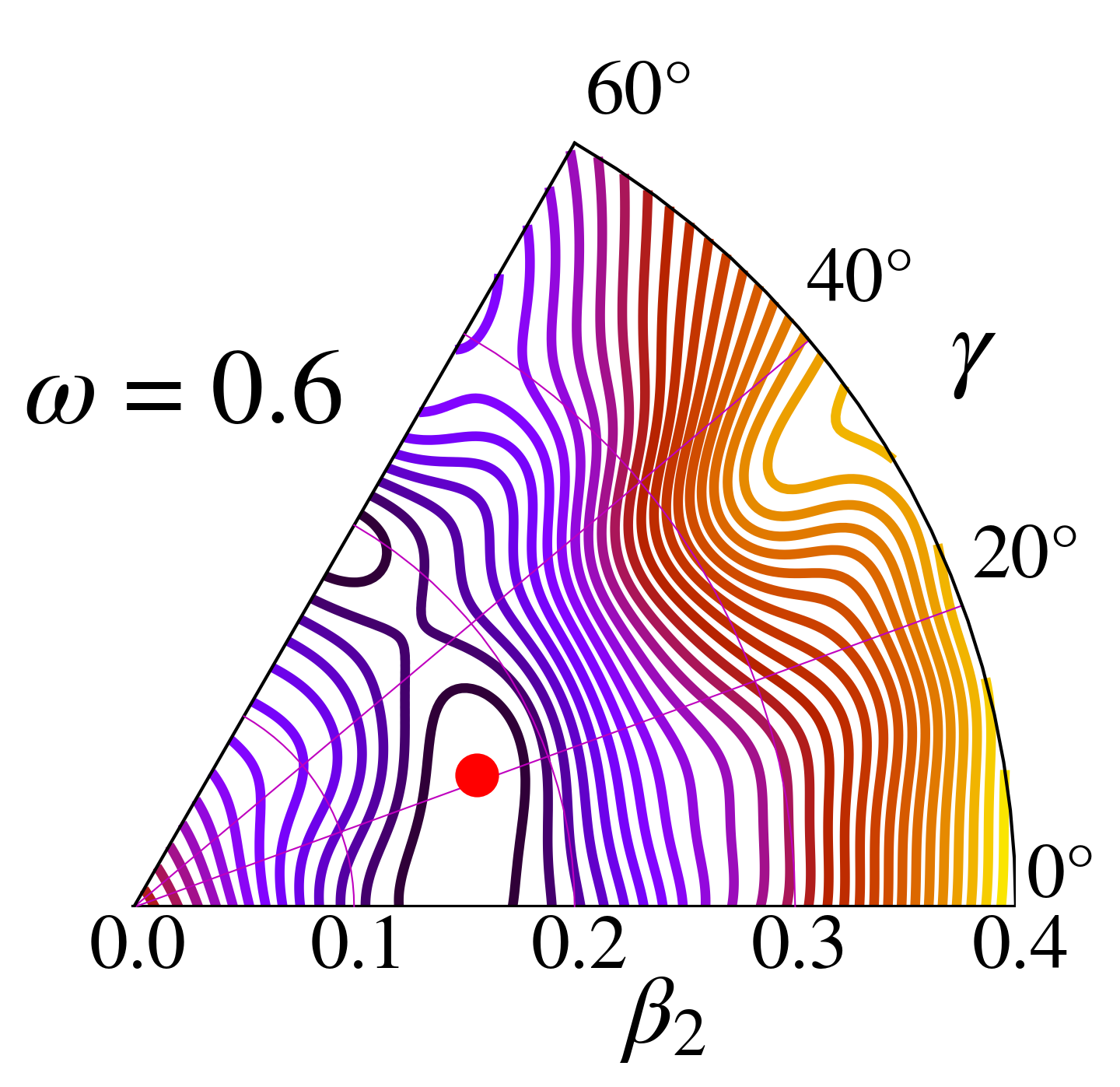}
\includegraphics[width=0.32\linewidth]{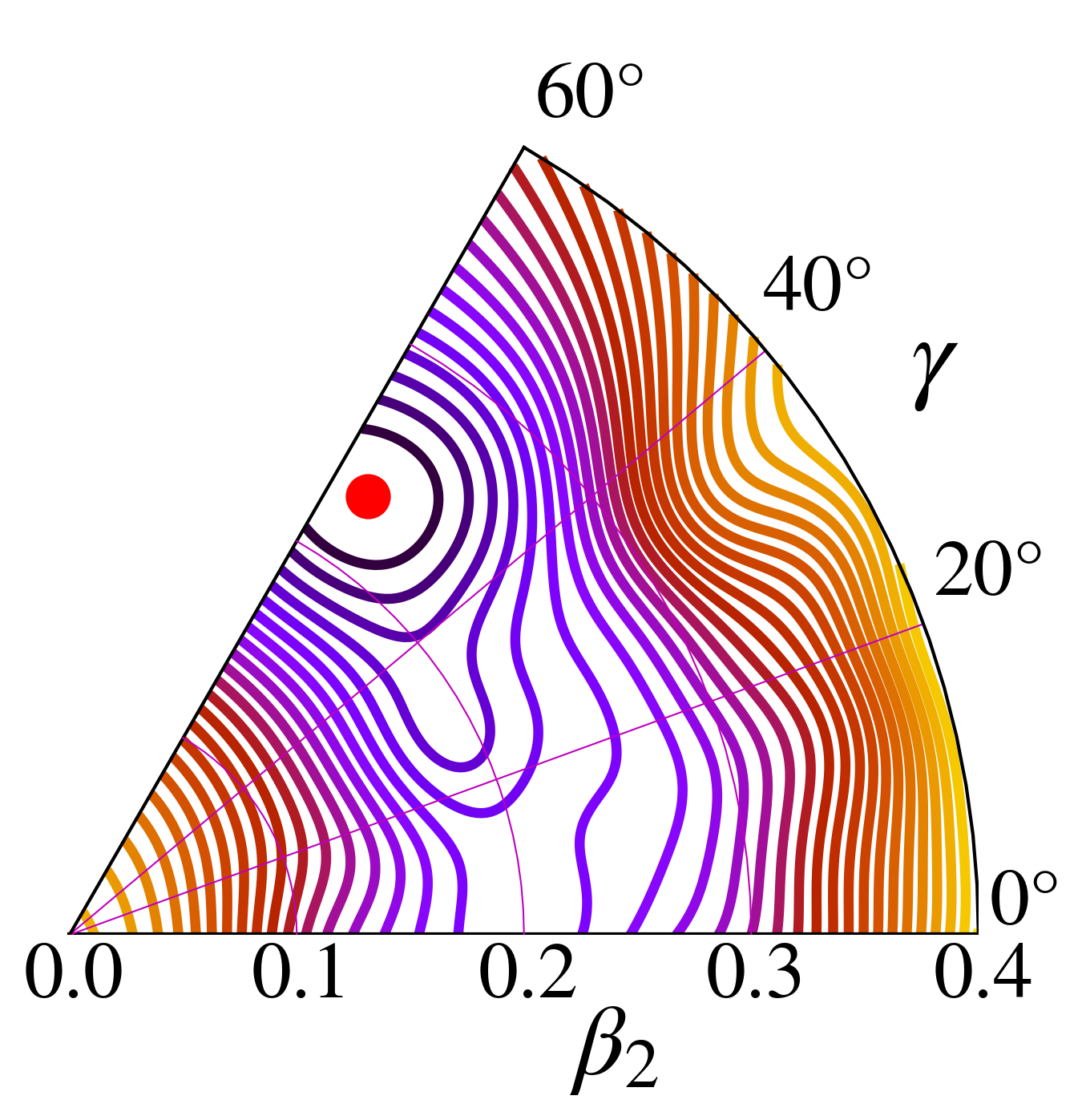}
\includegraphics[width=0.32\linewidth]{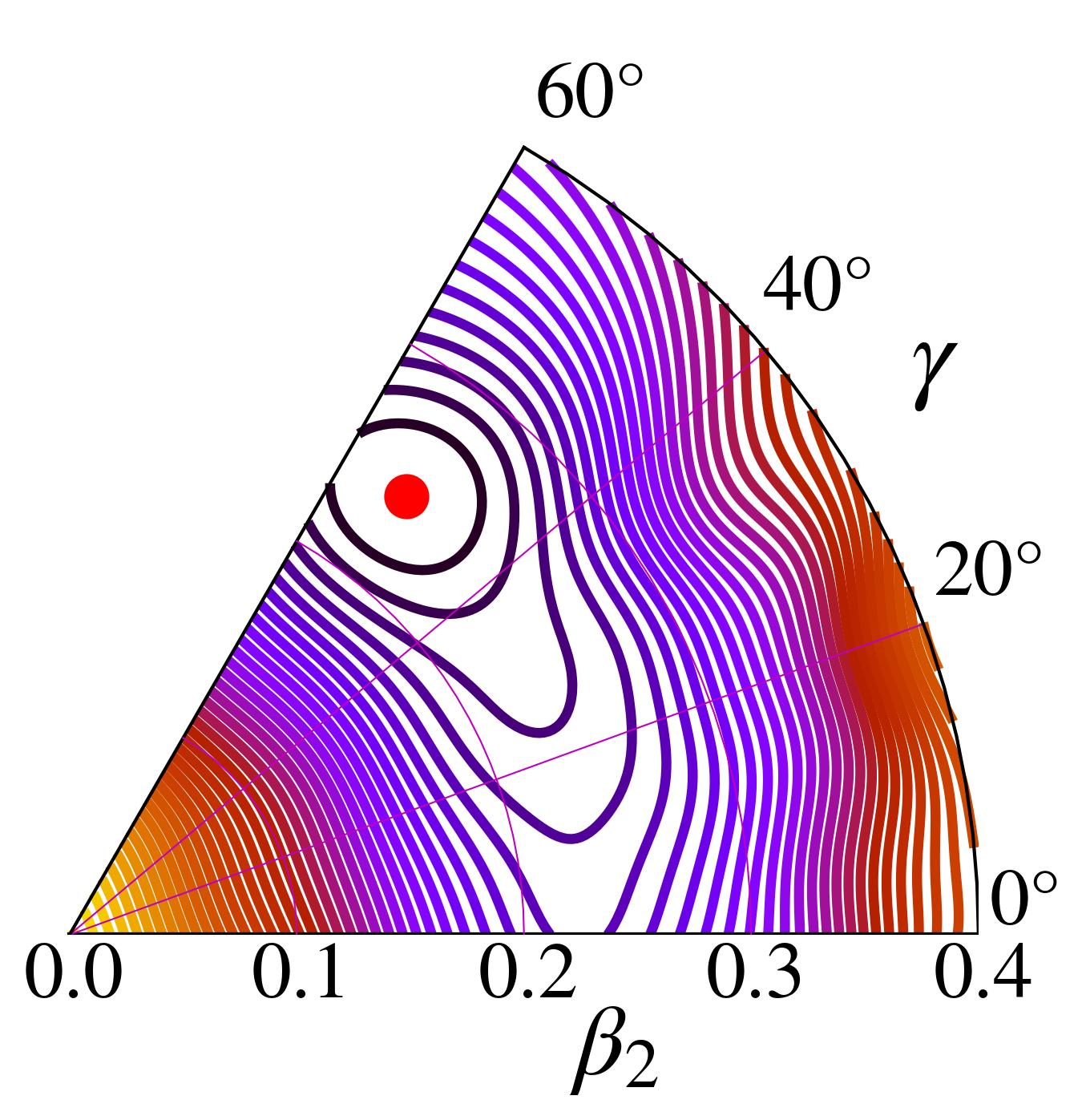}

\caption{Total Routhian surface plots at four different rotational frequencies, $\omega = 0.0$, 0.2, 0.4, and 0.6 (MeV/$\hbar$) in the `a' configuration for $^{87,89,91}$Br isotopes. The red-filled circle represents deformation at TRS minimum, and the contour line spacing is $0.2$ MeV. Each row corresponds to a fixed rotational frequency, while the columns correspond to the three isotopes.}
\label{TRS_A}
\end{figure}

Figure~\ref{TRS_A} shows the TRS contour plots for $^{87,89,91}$Br at rotational frequencies $\omega = 0.0$, 0.2, 0.4, and 0.6 MeV/$\hbar$. The corresponding quadrupole deformation $\beta_2$ and triaxiality parameter $\gamma$ values at the energy minima are summarized in Table~\ref{tab:tab3}.

\begin{table}[h!]
\caption{\label{tab:tab3}The quadrupole $(\beta_2)$ and triaxial $(\gamma)$ deformation parameters of $^{87,89,91}$Br for configuration `a' at different angular frequency $\omega$.}

\begin{ruledtabular}
\begin{tabular}{lcccc}
Nuclei & $\omega = 0.0$ & $\omega = 0.2$ & $\omega = 0.4$ & $\omega = 0.6$ \\
\colrule
\\[-0.6em]
$^{87}$Br & 0.1668 & 0.1668 & 0.1668 & 0.1668 \\
          & $0^\circ$--$35^\circ$ & $4^\circ$--$38^\circ$ & $0^\circ$--$35^\circ$ & $0^\circ$--$35^\circ$ \\
\hline
\\[-0.6em]
$^{89}$Br & 0.1768 & 0.1768 & 0.1768 & 0.2330 \\
          & $10^\circ$--$35^\circ$ & $10^\circ$--$35^\circ$ & $10^\circ$--$35^\circ$ & $45^\circ$--$60^\circ$ \\
\hline
\\[-0.6em]
$^{91}$Br & 0.2190 & 0.2190 & 0.2090 & 0.2432 \\
          & $45^\circ$--$60^\circ$ & $45^\circ$--$60^\circ$ & $0^\circ$--$35^\circ$ & $45^\circ$--$60^\circ$ \\

\end{tabular}
\end{ruledtabular}
\end{table}

For $^{87}$Br, the energy minimum remains relatively stable with $\beta_2 \approx 0.17$ over the entire frequency range, indicating a moderately deformed shape. The $\gamma$ values span from near-axial to weakly triaxial configurations ($0^\circ$–$38^\circ$), suggesting $\gamma$-soft behavior with no pronounced shape transition as a function of rotational frequency.

In contrast, $^{89}$Br exhibits a noticeable evolution of shape with increasing rotational frequency. While the nucleus maintains a moderate quadrupole deformation ($\beta_2 \approx 0.18$) at low frequencies, a transition toward larger deformation and pronounced triaxiality is observed at $\omega = 0.6$ MeV/$\hbar$, where the minimum shifts to $\beta_2 \approx 0.23$ and $\gamma \approx 45^\circ$–$60^\circ$. This behavior indicates the onset of a rotation-induced shape change driven by quasiparticle alignment effects.

For $^{91}$Br, the TRS minima display a more complex evolution. At lower rotational frequencies, the nucleus favors a triaxial shape with $\gamma \approx 45^\circ$–$60^\circ$ and $\beta_2 \approx 0.22$. At $\omega = 0.4$ MeV/$\hbar$, the minimum shifts toward a near-axial shape with a slightly reduced deformation, followed by a return to a more strongly triaxial and enhanced deformation ($\beta_2 \approx 0.24$) at $\omega = 0.6$ MeV/$\hbar$. This behavior reflects a delicate interplay between shell structure, pairing correlations, and rotational alignment in neutron-rich Br isotopes.

Overall, the configuration-constrained TRS analysis reveals that while $^{87}$Br remains $\gamma$-soft with stable deformation, $^{89}$Br and $^{91}$Br exhibit clear rotation-driven shape evolution, including enhanced triaxiality and deformation at higher rotational frequencies. These results provide important microscopic insight into the observed rotational behavior and support the interpretation of quasiparticle alignment effects in these nuclei.

\end{document}